\documentclass[english,aps, prl, twocolumn]{revtex4-2}
\usepackage[T1]{fontenc}
\usepackage[utf8]{inputenc}
\usepackage{amsmath}
\usepackage{braket}
\usepackage[pdftex]{graphicx} 		   
\usepackage[percent]{overpic}
\usepackage{tikz}
\usepackage{tikz-feynman}
\usetikzlibrary{patterns,arrows.meta}

\usepackage{babel}

\begin{document}

\title{Self-organized criticality in complex model ecosystems}
\author{Thibaut Arnoulx de Pirey}
\email{thibaut.arnoulxdepirey@ipht.fr}
\affiliation{Institut de Physique Th\'eorique, CEA Saclay \& CNRS (UMR 3681), Gif-sur-Yvette,
91190, France}

\begin{abstract}
We show that spatial extensions of many-species population dynamics
models, such as the Lotka-Volterra model with random interactions
we focus on in this work, generically exhibit scale-free correlation
functions of population sizes in the limit of an infinite number of
species. Using dynamical mean-field theory, we describe the many-species
system in terms of single-species dynamics with demographic and environmental
noises. We show that the single-species model features a random mass
term, or equivalently a random space-time averaged growth rate, poising
some species very close to extinction. This introduces a hierarchy
of ever larger correlation times and lengths as the extinction threshold
is approached. In turn, every species, even those far from extinction,
are coupled to these near-critical fields which combine to make fluctuations
of population sizes generically scale-free. We argue that these correlations
are described by exponents derived from those of directed percolation
in spatial dimension $d=3$, but not in lower dimensions.
\end{abstract}

\maketitle

A central result in theoretical ecology, the competitive exclusion
principle \citep{levin1970community,mittelbach2019community}, states
that an ecosystem at equilibrium cannot sustain more species than
the number of available resources or niches---that is, partitions
of the habitat within which a species can survive and reproduce. Yet
many natural ecosystems exhibit a staggering diversity of species
thriving on a seemingly limited number of resources and possible niches,
an apparent contradiction termed the ``paradox of the plankton''
\citep{hutchinson1961paradox}. The commonly accepted resolution is
that planktonic communities do not reach an equilibrium for which
only a few species would prevail \citep{scheffer2003plankton}, and
numerous mechanisms have been proposed to explain the persistence
of these nonequilibrium states. At fine taxonomic resolution, plankton
populations are indeed known to undergo large, erratic fluctuations
with species turnover \citep{martin2018high}, meaning continuous
changes in the identity of the most abundant species. That these fluctuations can persist over several years under homogeneous and constant
environmental conditions \citep{beninca2008chaos} highlights the
plausibly endogenous origin of such dynamics. More generally,
population fluctuations are widespread in natural ecosystems \citep{rogers2022chaos},
such as in terrestrial bacterial communities, both over time \citep{grilli2020macroecological}
and space \citep{louca2016high}. 

At the theoretical level, there has been substantial interest in recent
years in modeling highly diverse ecosystems, with the goal of understanding
the typical behaviors that emerge from the interactions of many species
with heterogeneous traits and strategies. A robust conclusion from
these studies is that spatially homogeneous ecosystems with weak variability
in the interspecific couplings tend to converge, at long times, to
an equilibrium, whereas those with stronger heterogeneity exhibit
persistent population fluctuations, even in the absence of environmental
variation. These two regimes are separated by a sharp phase transition
in the limit of a large number of species. Building upon the seminal
work \citep{may1972will} of R. May on the stability of fixed points
in random dynamical systems, similar conclusions have been obtained
for classical models of population dynamics, including the replicator
equation \citep{opper1992phase}, the Lotka-Volterra model \citep{bunin2017ecological}
and the Mac-Arthur resource-consumer model \citep{blumenthal2024phase}.
In that sense, nonequilibrium states are typical outcomes of many-species
dynamics. In some cases, when the existence of a Lyapunov function
is enforced by a symmetry in the ensemble of interspecific interactions,
the fluctuating phase is replaced by a coexistence of multiple marginal
equilibria \citep{biroli2018marginally}. Remarkably, experiments
on randomly assembled bacterial communities display a similar transition
between a fixed-point regime and a fluctuating one \citep{hu2022emergent}.

However, for spatially homogeneous populations, the resulting nonequilibrium
states are generically pathological and cannot offer a satisfactory
resolution to the paradox of the plankton as they exhibit aging---or dynamical slowdown \citep{arnoulx2024many,arnoulx2023aging,pearce2020stabilization,roy2019numerical}.
This occurs because the deterministic dynamics drives the population
size of any species arbitrarily close to zero, provided one waits
long enough. Although populations eventually recover from these dips
at the deterministic level, introducing even a small amount of demographic
noise (square-root noise that can cause finite-time extinctions) or
imposing a deterministic extinction threshold below which recovery
is impossible leads to the rapid extinction of many species and to
the suppression of the fluctuating phase altogether \citep{roy2020complex,pearce2020stabilization}.
At the deterministic level, this issue can be mitigated by introducing
a finite migration rate. The underlying assumption is that the ecosystem
under study is coupled to a much larger environment in which species
diversity is maintained, and from which a constant flux of individuals
maintains local populations away from extinction. When the migration
rate is small, the long-time dynamics is time-translation invariant
and exhibits large fluctuations of population sizes with species turnover
\citep{arnoulx2024many,mallmin2024chaotic}. These results, however,
do not address how diversity is sustained in the surrounding environment.

A natural extension is therefore to go beyond the well-mixed assumption
and investigate the spatio-temporal dynamics of such complex ecosystems.
In the metacommunity framework \citep{leibold2004metacommunity},
species interact locally with dispersion of populations between different
spatial locations. In \citep{pearce2020stabilization}, it was shown
that deterministic spatio-temporal chaos can be stabilized on an infinite,
fully connected spatial lattice, yielding a much higher overall diversity
than would be attainable in an equivalent well-mixed system. In the
same spatial setting, but with demographic noise, \citep{garcia2024interactions}
demonstrated the existence of a high-diversity phase even in parameter
regimes where each species, considered in isolation from the others,
would go extinct. They also found population-size distributions qualitatively
similar to those obtained in a well-mixed ecosystem with a fixed migration
rate. In these studies, desynchronized spatial fluctuations enable
the maintenance of diversity: fluctuations -- whether deterministic
or stochastic---may drive the local abundance of a species to zero,
but migration from locations where the species persists prevents global
extinction. This mechanism, akin to what is called a ``storage effect''
in ecology \citep{loreau2003biodiversity}, is generically at play
in nonequilibrium phase transitions between active and absorbing states
\citep{hinrichsen2000non}. 

In this article, we investigate the behavior of complex ecosystem
models embedded in a regular $d$-dimensional space. With short-range
(diffusive) dispersal, our main finding is that spatio-temporal correlations
of population sizes generically display scale-free behavior at large
scales. These fluctuations, arising from species interactions, are
therefore markedly different from the spatio-temporal fluctuations
experienced by a single population in a changing environment. In these
models, self-organized criticality arises as a consequence of the
limit of an infinite number of species. Importantly, the associated
exponents appear to define broad universality classes, in contrast
to the exponents governing the rare tail of species abundance distributions,
which can be strongly model- or parameter-dependent \citep{mallmin2024chaotic}.
This result reinforces the idea that studying spatio-temporal correlations
in ecological data may provide key insights into the dominant mechanisms
structuring highly diverse ecosystems. In spatial dimension $d=3$,
we conjecture that the exponents can be derived from those of directed
percolation. This is not the case in $d=1$ and $d=2$, however, as
species interactions generate fluctuations that are too long-range
to preserve the single-species directed percolation universality class.
Finally, we show that these results extend to the two-time correlation
function in the presence of very long-range dispersal---that is
on the fully connected lattice---although with an amplitude that
decreases with the number of lattice sites.

\emph{Dynamical mean-field theory}---We consider an ecosystem harboring
$S\gg1$ species in a homogeneous $d$-dimensional environment, with
local density at position $x$ and time $t$ denoted by $n_{i}(x,t)$
for $i=1\dots S$. We study the simplest spatial extension of the
Lotka-Volterra dynamics,

\begin{equation}
\tau\partial_{t}n_{i}=n_{i}\left(1-n_{i}-\sum_{j\neq i}\alpha_{ij}n_{j}\right)+D\Delta n_{i}+\sqrt{2Tn_{i}}\,\eta_{i}(x,t)\,,\label{eq:many_body_GLV}
\end{equation}

\noindent with randomly sampled interspecific weak couplings such
that $\left\langle \alpha_{ij}\right\rangle =\mu/S$ and $\left\langle \alpha_{ij}\alpha_{kl}\right\rangle =\delta_{ik}\delta_{jl}\sigma^{2}/S$.
The ${\eta_{i}}$ are independent zero-mean Gaussian space-time white
noises, with $\left\langle \eta_{i}(x,t)\eta_{j}(x',t')\right\rangle =\delta_{ij}\delta(t-t')\delta(x-x')$,
and account for the stochasticity of birth-death processes, see Chap.
12 of \citep{krapivsky2010kinetic}. The above equation is understood
as Ito-discretized. We assume that the amplitude $T$ of the demographic
noise is the same across all species. We further assume that all species
have the same carrying capacity (rescaled here to the value $1$ reached
in the steady state by the population sizes in the absence of noise
and interactions), generation time $\tau$, and diffusion coefficient
$D$ (both set to $1$ in the following), so that we focus on the
consequences of interspecific interactions. A straightforward extension
of dynamical mean-field theory \citep{sompolinsky1982relaxational}
built for the $0$-dimensional version of the Lotka-Volterra dynamics
\citep{roy2019numerical,galla2024generating} shows that, as $S\to\infty$,
the population size of any species evolves according to the noisy
Fisher--KPP equation
\begin{equation}
\partial_{t}n=n(g(x,t)-n)+\Delta n+\sqrt{2Tn}\,\eta(x,t)\,.\label{eq:seascape}
\end{equation}

\noindent The growth rate $g(x,t)$, originating from the interactions
with all the other species, is a Gaussian process whose correlations
obey the following self-consistency conditions
\[
\left\langle g(x,t)\right\rangle =1-\mu\left\langle n(x,t)\right\rangle \,,
\]
\[
\left\langle g(x,t)g(x',t')\right\rangle _{c}=\sigma^{2}\left\langle n(x,t)n(x',t')\right\rangle \,.
\]

\noindent When the noise $g(x,t)$---often called environmental
or seascape noise---has short-range correlations, similar dynamics
to the one written in Eq.~(\ref{eq:seascape}) have been studied
\citep{grinstein1997statistical,ottino2020population,van1998wilson},
and connected to the directed percolation \citep{hinrichsen2000non}
and Kardar-Parisi-Zhang (KPZ) \citep{kardar1986dynamic} universality
classes, when $T>0$ and $T=0$ respectively. 

\emph{Emergence of scale-free correlations---}Here the correlations
of the noise $g(x,t)$ are not given in advance and must be determined
self-consistently. First, we note that in the absence of interactions,
$\sigma=\mu=0$, we have $g(x,t)=1$. In that case, in any dimension
$d\geq1$, there exists a critical $T_{c}$ such that the dynamics
in Eq.~(\ref{eq:seascape}) exhibits an extinct phase $n(x,t)=0$
at high noise $T>T_{c}$, and an active one $\left\langle n\right\rangle >0$
for $T<T_{c}$. We consider $T<T_{c}$ in the following, ensuring
that a finite fraction of species survives in the steady state even
in the presence of interactions. We further assume that a time- and
space-translation invariant steady state is eventually reached. Because
population sizes are positive $\left\langle n\right\rangle >0$, we
can decompose \citep{arnoulx2025critical,garcia2024interactions}
the noise $g(x,t)$ as $g(x,t)=\bar{g}+\delta g(x,t)$ where $\bar{g}$
is a frozen homogeneous Gaussian random variable and $\delta g(x,t)$
is an independent zero-mean fluctuating contribution that decorrelates
over long times, meaning $\left\langle \delta g(x,t)\delta g(x',t')\right\rangle \to0$
as $|t-t'|\to\infty$. The resulting self-consistency equations become
\[
\left\langle \bar{g}\right\rangle =1-\mu\int_{-\infty}^{+\infty}d\bar{g}\,P(\bar{g})\left\langle n\right\rangle _{\bar{g}}\,,
\]
\begin{equation}
\left\langle \bar{g}^{2}\right\rangle _{c}=\sigma^{2}\int_{-\infty}^{+\infty}d\bar{g}\,P(\bar{g})\left\langle n\right\rangle _{\bar{g}}^{2}\,,\label{eq:variance}
\end{equation}
and

\begin{widetext} \begin{equation} \label{eq:self_consistency} \left\langle \delta g(x,t)\delta g(x',t')\right\rangle=\sigma^{2}\int_{-\infty}^{+\infty}d\bar{g}\,P(\bar{g})\left\langle \left(n(x,t)-\left\langle n\right\rangle _{\bar{g}}\right)\left(n(x',t')-\left\langle n\right\rangle _{\bar{g}}\right)\right\rangle _{\bar{g}}\,,\end{equation} \end{widetext}

\noindent where the averages $\left\langle \dots\right\rangle _{\bar{g}}$
are taken over the stochastic process
\begin{equation}
\partial_{t}n=n(\bar{g}+\delta g(x,t)-n)+D\Delta n+\sqrt{2Tn}\,\eta(x,t)\,,\label{eq:field_theory_bis}
\end{equation}

\noindent at a fixed value of $\bar{g}$. The space-time averaged
growth rate $\bar{g}$ has a finite variance across species because
a finite fraction of them survives in the steady state, see Eq.~(\ref{eq:variance}).

The self-consistency equations of the dynamical mean-field theory
must be solved at all scales and are therefore generically not tractable.
We can show, however, that space-time translation invariant solutions
must exhibit a scale-free behavior at large scales. This arises because
the dynamics in Eq.~(\ref{eq:field_theory_bis}) features a random
mass term $\propto n\bar{g}$. The latter is generated by species
interactions because \textit{(i)} the variables $n$ are positive
so that the self-consistent noise acquires a frozen part $\bar{g}$
and \textit{(ii)} the growth is multiplicative which makes this frozen
part appear as a mass term. Assuming short-range $\delta g(x,t)$
in Eq.~(\ref{eq:field_theory_bis}), there exists a critical value
of $\bar{g}$ such that species go extinct $\left\langle n\right\rangle _{\bar{g}}=0$
for $\bar{g}<\bar{g}_{c}$, while they survive for $\bar{g}>\bar{g}_{c}$.
The corresponding critical point lies in the directed percolation
universality class, because environmental fluctuations are negligible
compared to demographic ones when $n\ll1$. Close to the extinction
threshold, $0<\bar{g}-\bar{g}_{c}\ll1$, species evolve over time
and length scales that become arbitrarily large as $|\bar{g}-\bar{g}_{c}|\to0$.
As these species affect the dynamics of all the others through the
feedback loop expressed by the self-consistency conditions, the system
become scale-free in the limit of an infinite number of species where
$\min_{{\rm species}}(|\bar{g}-\bar{g}_{c}|)\to0$. More precisely,
in any dimension $d<4$ (with $d=4$ being the upper critical dimension
of directed percolation), and for short-range $\delta g(x,t)$, near-critical
correlations are described by the scaling function 
\[
\left\langle \left(n(x,t)-\left\langle n\right\rangle _{\bar{g}}\right)\left(n(0,0)-\left\langle n\right\rangle _{\bar{g}}\right)\right\rangle _{\bar{g}}=x^{-2\zeta}\mathcal{F}\left(\frac{x}{\xi},\frac{t}{\tau}\right)\,,
\]

\noindent where the correlation length $\xi\sim|\bar{g}-\bar{g}_{c}|^{-\nu}$
and correlation time $\tau\sim|\bar{g}-\bar{g}_{c}|^{-z\nu}$ and
with directed percolation exponents $\zeta,\,\nu$ and $z$---$\zeta$
being related to the order parameter exponent $\beta$ given by $\left\langle n\right\rangle _{\bar{g}}\sim|\bar{g}-\bar{g}_{c}|^{\beta}$
through $\beta=\zeta\nu$. After integrating over $\bar{g}$, the
large-scale result is dominated by the vicinity of the critical point
and we obtain the scale-free behavior
\begin{align}
\int_{-\infty}^{+\infty}d\bar{g}\,P(\bar{g})\left\langle \left(n(x,t)-\left\langle n\right\rangle _{\bar{g}}\right)\left(n(0,0)-\left\langle n\right\rangle _{\bar{g}}\right)\right\rangle _{\bar{g}}\nonumber \\
\sim P(\bar{g}_{c})x^{-(2\beta+1)/\nu}\mathcal{G}\left(tx^{-z}\right) & \,,\label{eq:scale-free}
\end{align}

\noindent where the scaling function $\mathcal{G}$ is given by 
\[
\mathcal{G}(y)=\int_{0}^{\infty}du\,\mathcal{F}\left(u^{\nu},yu^{z\nu}\right)\,.
\]

\noindent As mentioned, $\bar{g}$ is a Gaussian variable with a positive
variance, and thus $P(\bar{g}_{c})\neq0$. Therefore, there are no
solution of Eq.~(\ref{eq:self_consistency}) with exponential-like
correlations: The fluctuations in population sizes are necessarily
scale-free. In a system with a finite number of species $S\gg1$,
we expect instead $\min_{{\rm species}}(|\bar{g}-\bar{g}_{c}|)\sim S^{-1}$
and the system to exhibit a correlation length $\xi_{S}$ and correlation
time $\tau_{S}$ that diverge with $S$ as $\xi_{S}\sim S^{\nu}$
and $\tau_{S}\sim S^{z\nu}$. The same argument shows that there is
no short-range solution of the dynamical mean-field theory equation
in the case $T=0$ where the underlying many-species dynamics is deterministic.
Interestingly, the existence of long-time power-law tails was already
mentioned \citep{pearce2020stabilization} in a deterministic ($T=0$)
and well-mixed population dynamics model. This required perfectly
antisymmetric interactions ($\alpha_{ij}=-\alpha_{ji}$), a case for
which the aging regime discussed in the introduction does not apply.
These tails were traced back to the presence of species with very
small averaged growth rates, see appendix 3.D. therein. 

\emph{Determining the critical exponents---}We have shown that an
effective environmental noise $\delta g(x,t)$ with exponential-like
space-time correlations yields species-averaged fluctuations
of population sizes with a scale-free form given in Eq.~(\ref{eq:scale-free}).
Naturally, the next step is to assume that the effective environmental
noise has large-scale correlations given by Eq.~(\ref{eq:scale-free})
and to look at the resulting fluctuations of population sizes. We
expect them to be scale-free, but are the associated exponents the
same? To answer this question, we consider the dynamics in Eq.~(\ref{eq:field_theory_bis})
with a long-range correlated noise $\delta g(x,t)$ 
\[
\left\langle \delta g(x,t)\delta g(0,0)\right\rangle =x^{-2\mu}\tilde{\mathcal{G}}\left(tx^{-z}\right)\,,
\]

\noindent where the exponent $\mu$ is arbitrary and $z$ is the dynamical
critical exponent of the directed percolation universality class in
the same dimension. Building upon the study of the KPZ equation with
long-range correlated noise \citep{medina1989burgers,janssen1999exact},
we find that, to linear order in the amplitude of the environmental
noise, the directed percolation universality class is preserved as
long as $\mu>d+z-2\zeta$, see \citep{supp}. Setting $2\mu=(2\beta+1)/\nu$
and using numerical estimates of the exponents $\beta,\,\nu$ and
$z$ \citep{hinrichsen2000non} shows that the directed percolation
scaling is preserved in $d=3$ but not in $d=1$ and $d=2$.
We thus conjecture that the scaling form in Eq.~(\ref{eq:scale-free})
correctly describes the large-scale population fluctuations in $d=3$---at least in a regime of small $\sigma$, where the amplitude of
the environmental noise is small, see Eq.~(\ref{eq:self_consistency}).
Similar conclusions should also hold for $\sigma<\sigma_{c}$ and
small $T$, where $\sigma_{c}=\sqrt{2}$ is the critical value below
which the well-mixed $T=0$ dynamics converges to a fixed point \citep{bunin2017ecological}.
This yields the $d=3$ two-point correlation function
\[
\left\langle \delta g(x,t)\delta g(x',t)\right\rangle \sim|x-x'|^{-(2\beta+1)/\nu}\sim|x-x'|^{-4.5}\,,
\]

\noindent and the two-time correlation function
\[
\left\langle \delta g(x,t)\delta g(x,t')\right\rangle \sim|t-t'|^{-(2\beta+1)/z\nu}\sim|t-t'|^{-2.4}\,.
\]
In lower dimension, the decay exponents cannot be simply derived from those of directed
percolation and are yet to be determined. From an ecological point
of view, it is important to note that spatial and temporal scales
are intertwined in the scaling form of Eq.~(\ref{eq:scale-free}). 

We conclude by extending these results above the upper critical dimension
$d>4$, and on the fully connected lattice. In $d>4$, and in the
presence of short-range environmental noise $\delta g(x,t)$, the
large-scale near-critical physics of Eq.~(\ref{eq:field_theory_bis})
is described by a weak-noise Gaussian fixed point. After integrating
over the fluctuating mass term $\bar{g}$, see \citep{supp}, we get
\begin{align}
\int_{-\infty}^{+\infty}d\bar{g}\,P(\bar{g})\left\langle \left(n(x,t)-\left\langle n\right\rangle _{\bar{g}}\right)\left(n(0,0)-\left\langle n\right\rangle _{\bar{g}}\right)\right\rangle _{\bar{g}}\nonumber \\
\sim x^{-(d+2)}\mathcal{G}\left(tx^{-2}\right)\,.\label{eq:correl_large_d}
\end{align}

\noindent Importantly, the correlation function on the right-hand
side of Eq.~(\ref{eq:correl_large_d}) is space-time integrable in
any $d>4$. We therefore conjecture that it correctly describes the
solution of the dynamical mean-field theory equation at large scales.
We lastly discuss the case of the fully connected network with $M\gg1$
sites. As in finite dimension, the extension of Eq.~(\ref{eq:field_theory_bis})
with short-range environmental noise to the fully connected lattice
exhibits a phase transition between an extinct and an active phase
as $M\to\infty$ \citep{ottino2020population}. However, critical
slowing down only emerges at the level of the small $O(1/\sqrt{M})$
Gaussian fluctuations of the spatially averaged population sizes.
This leads to the emergence of power-law tails, although with an amplitude
that decreases with $M$, see \citep{supp}. More precisely, using
Greek indices to label the different sites of the lattice, we expect
the solution of the dynamical mean-field theory equation to behave
as $\left\langle \delta g_{\mu}(t)\delta g_{\nu}(t')\right\rangle \sim A_{M}/|t-t'|$
at large $|t-t'|$ (but $|t-t'|$ small compared to $\sqrt{M}$ due
to finite-time extinctions, when $M$ is finite, of species that would
survive but be on average rare when $M\to\infty$), with $A_{M}\sim M^{-1}$,
and up to logarithmic-in-time corrections.

\medskip{}

In this work, we have shown that complex model ecosystems with disordered
interactions generically exhibit scale-free spatio-temporal correlations,
in which spatial and temporal scales are strongly intertwined. These
correlations are characterized by exponents that are expected to be
universal. The scale-free structure arises because interspecific interactions
drive some species to the brink of extinction, as reflected in the
Gaussian distribution of space-time averaged growth rates across species.
The dynamics of these species, in turn, exhibit increasingly large
correlation times and lengths, which feed back -- even though these
species are rare -- into the the entire system. We expect this mechanism
to be present in other disordered, high-dimensional population dynamics
models. Interestingly, rarity is common in ecology \citep{lynch2015ecology}.
This work highlights the important role that species that are rare
on average may play on ecosystem dynamics, provided that these dynamics
are at least partially driven by species interactions. 

It is instructive to compare the present work with \citep{denk2022self},
which identified a different mechanism for self-organized criticality
in many-species ecosystems. In that study, the authors considered
the Lotka-Volterra model with homogeneous competitive interactions.
Due to the overall competition, the population size of each species
decreases as the number of species increases. In the limit of infinitely
many species -- which we expect to correspond to $\sigma=0$ and
$\mu\to\infty$ in the model studied here -- each population approaches
extinction, giving rise to scale-free fluctuations characterized by
exponents consistent with the directed percolation universality class.
In contrast, when interactions are heterogeneous ($\sigma>0$), the
phenomenology changes: not all species have small average abundances,
and the resulting correlations differ from those of a single field
near the directed percolation transition. 

We conclude this work by some perspectives. First, it would be very
interesting to explore the ecological consequences of these long-range
correlations. While the distribution of averaged growth rates leads
to the extinction of a finite fraction of species, as observed on
the fully connected lattice \citep{garcia2024interactions}, the exact
fraction and the impact of these scale-free fluctuations on diversity
remain unknown. We have reasons to believe that this might be a subtle
issue: in the deterministic well-mixed dynamics \citep{arnoulx2024many},
the infinitely long aging correlation time is both what makes every
species vulnerable (each population size can get arbitrarily close
to $0$) and what allows every species to compensate the influence
of the frozen part $\bar{g}$ of their growth rate---eventually allowing
every species to survive and making the system closer to neutrality
than one might expect. It would also be interesting to extend these
results to the deterministic case ($T=0$). As discussed here, no
time-translation invariant regime---unless the system reaches a fixed
point---can be described by exponential-like correlations. However, whether a time-translation invariant state exists in
finite dimension remains unclear as the dynamics may synchronize across
the entire space---leading the ecosystem to behave as a single well-mixed
community. Indeed, because well-mixed dynamics exhibit aging and no
fast time scale, we expect the Lyapunov exponent of the zero-dimensional
system to vanish, making the synchronized state linearly stable in
any finite-size system. Numerics point to the absence of synchronization
on the infinite fully connected lattice \citep{pearce2020stabilization,roy2020complex},
but no results are available in finite dimension. Finally, it would
be interesting to investigate how these results are affected by frozen
spatial heterogeneity in the environment. Following \citep{garcia2024interactions,roy2020complex},
such heterogeneity could be modeled by a space-dependent interaction
matrix, $\left\langle \alpha_{ij}(x)\alpha_{ij}(x')\right\rangle =S^{-1}(\sigma^{2}+C(x-x'))$,
with $C(x)$ short-ranged. Because the directed percolation universality
class is destroyed in any dimension in the presence of frozen disorder
\citep{janssen1997renormalized,hooyberghs2004absorbing,noest1986new},
we expect that the results presented in this work would also be modified. 

\medskip{}
\medskip{}

\begin{acknowledgments}
I thank Giulia Garcia Lorenzana and Fr\'ed\'eric van Wijland for valuable
discussions. This work was partially conducted at the Center for Interdisciplinary
Research (ZiF) of the Bielefeld University during the program ``The
Lush World of Random Matrices''. I thank the organizers of the program
and the ZiF for the kind hospitality. 
\end{acknowledgments}

\noindent \bibliographystyle{apsrev4-2}
\bibliography{biblio}

\begin{thebibliography}{45}%
\makeatletter
\providecommand \@ifxundefined [1]{%
 \@ifx{#1\undefined}
}%
\providecommand \@ifnum [1]{%
 \ifnum #1\expandafter \@firstoftwo
 \else \expandafter \@secondoftwo
 \fi
}%
\providecommand \@ifx [1]{%
 \ifx #1\expandafter \@firstoftwo
 \else \expandafter \@secondoftwo
 \fi
}%
\providecommand \natexlab [1]{#1}%
\providecommand \enquote  [1]{``#1''}%
\providecommand \bibnamefont  [1]{#1}%
\providecommand \bibfnamefont [1]{#1}%
\providecommand \citenamefont [1]{#1}%
\providecommand \href@noop [0]{\@secondoftwo}%
\providecommand \href [0]{\begingroup \@sanitize@url \@href}%
\providecommand \@href[1]{\@@startlink{#1}\@@href}%
\providecommand \@@href[1]{\endgroup#1\@@endlink}%
\providecommand \@sanitize@url [0]{\catcode `\\12\catcode `\$12\catcode
  `\&12\catcode `\#12\catcode `\^12\catcode `\_12\catcode `\%12\relax}%
\providecommand \@@startlink[1]{}%
\providecommand \@@endlink[0]{}%
\providecommand \url  [0]{\begingroup\@sanitize@url \@url }%
\providecommand \@url [1]{\endgroup\@href {#1}{\urlprefix }}%
\providecommand \urlprefix  [0]{URL }%
\providecommand \Eprint [0]{\href }%
\providecommand \doibase [0]{https://doi.org/}%
\providecommand \selectlanguage [0]{\@gobble}%
\providecommand \bibinfo  [0]{\@secondoftwo}%
\providecommand \bibfield  [0]{\@secondoftwo}%
\providecommand \translation [1]{[#1]}%
\providecommand \BibitemOpen [0]{}%
\providecommand \bibitemStop [0]{}%
\providecommand \bibitemNoStop [0]{.\EOS\space}%
\providecommand \EOS [0]{\spacefactor3000\relax}%
\providecommand \BibitemShut  [1]{\csname bibitem#1\endcsname}%
\let\auto@bib@innerbib\@empty
\bibitem [{\citenamefont {Levin}(1970)}]{levin1970community}%
  \BibitemOpen
  \bibfield  {author} {\bibinfo {author} {\bibfnamefont {S.~A.}\ \bibnamefont
  {Levin}},\ }\href@noop {} {\bibfield  {journal} {\bibinfo  {journal} {The
  American Naturalist}\ }\textbf {\bibinfo {volume} {104}},\ \bibinfo {pages}
  {413} (\bibinfo {year} {1970})}\BibitemShut {NoStop}%
\bibitem [{\citenamefont {Mittelbach}\ and\ \citenamefont
  {McGill}(2019)}]{mittelbach2019community}%
  \BibitemOpen
  \bibfield  {author} {\bibinfo {author} {\bibfnamefont {G.~G.}\ \bibnamefont
  {Mittelbach}}\ and\ \bibinfo {author} {\bibfnamefont {B.~J.}\ \bibnamefont
  {McGill}},\ }\href@noop {} {\emph {\bibinfo {title} {Community ecology}}}\
  (\bibinfo  {publisher} {Oxford University Press},\ \bibinfo {year}
  {2019})\BibitemShut {NoStop}%
\bibitem [{\citenamefont {Hutchinson}(1961)}]{hutchinson1961paradox}%
  \BibitemOpen
  \bibfield  {author} {\bibinfo {author} {\bibfnamefont {G.~E.}\ \bibnamefont
  {Hutchinson}},\ }\href@noop {} {\bibfield  {journal} {\bibinfo  {journal}
  {The American Naturalist}\ }\textbf {\bibinfo {volume} {95}},\ \bibinfo
  {pages} {137} (\bibinfo {year} {1961})}\BibitemShut {NoStop}%
\bibitem [{\citenamefont {Scheffer}\ \emph {et~al.}(2003)\citenamefont
  {Scheffer}, \citenamefont {Rinaldi}, \citenamefont {Huisman},\ and\
  \citenamefont {Weissing}}]{scheffer2003plankton}%
  \BibitemOpen
  \bibfield  {author} {\bibinfo {author} {\bibfnamefont {M.}~\bibnamefont
  {Scheffer}}, \bibinfo {author} {\bibfnamefont {S.}~\bibnamefont {Rinaldi}},
  \bibinfo {author} {\bibfnamefont {J.}~\bibnamefont {Huisman}},\ and\ \bibinfo
  {author} {\bibfnamefont {F.~J.}\ \bibnamefont {Weissing}},\ }\href@noop {}
  {\bibfield  {journal} {\bibinfo  {journal} {Hydrobiologia}\ }\textbf
  {\bibinfo {volume} {491}},\ \bibinfo {pages} {9} (\bibinfo {year}
  {2003})}\BibitemShut {NoStop}%
\bibitem [{\citenamefont {Martin-Platero}\ \emph {et~al.}(2018)\citenamefont
  {Martin-Platero}, \citenamefont {Cleary}, \citenamefont {Kauffman},
  \citenamefont {Preheim}, \citenamefont {McGillicuddy}, \citenamefont {Alm},\
  and\ \citenamefont {Polz}}]{martin2018high}%
  \BibitemOpen
  \bibfield  {author} {\bibinfo {author} {\bibfnamefont {A.~M.}\ \bibnamefont
  {Martin-Platero}}, \bibinfo {author} {\bibfnamefont {B.}~\bibnamefont
  {Cleary}}, \bibinfo {author} {\bibfnamefont {K.}~\bibnamefont {Kauffman}},
  \bibinfo {author} {\bibfnamefont {S.~P.}\ \bibnamefont {Preheim}}, \bibinfo
  {author} {\bibfnamefont {D.~J.}\ \bibnamefont {McGillicuddy}}, \bibinfo
  {author} {\bibfnamefont {E.~J.}\ \bibnamefont {Alm}},\ and\ \bibinfo {author}
  {\bibfnamefont {M.~F.}\ \bibnamefont {Polz}},\ }\href@noop {} {\bibfield
  {journal} {\bibinfo  {journal} {Nature communications}\ }\textbf {\bibinfo
  {volume} {9}},\ \bibinfo {pages} {266} (\bibinfo {year} {2018})}\BibitemShut
  {NoStop}%
\bibitem [{\citenamefont {Beninc{\`a}}\ \emph {et~al.}(2008)\citenamefont
  {Beninc{\`a}}, \citenamefont {Huisman}, \citenamefont {Heerkloss},
  \citenamefont {J{\"o}hnk}, \citenamefont {Branco}, \citenamefont {Van~Nes},
  \citenamefont {Scheffer},\ and\ \citenamefont {Ellner}}]{beninca2008chaos}%
  \BibitemOpen
  \bibfield  {author} {\bibinfo {author} {\bibfnamefont {E.}~\bibnamefont
  {Beninc{\`a}}}, \bibinfo {author} {\bibfnamefont {J.}~\bibnamefont
  {Huisman}}, \bibinfo {author} {\bibfnamefont {R.}~\bibnamefont {Heerkloss}},
  \bibinfo {author} {\bibfnamefont {K.~D.}\ \bibnamefont {J{\"o}hnk}}, \bibinfo
  {author} {\bibfnamefont {P.}~\bibnamefont {Branco}}, \bibinfo {author}
  {\bibfnamefont {E.~H.}\ \bibnamefont {Van~Nes}}, \bibinfo {author}
  {\bibfnamefont {M.}~\bibnamefont {Scheffer}},\ and\ \bibinfo {author}
  {\bibfnamefont {S.~P.}\ \bibnamefont {Ellner}},\ }\href@noop {} {\bibfield
  {journal} {\bibinfo  {journal} {Nature}\ }\textbf {\bibinfo {volume} {451}},\
  \bibinfo {pages} {822} (\bibinfo {year} {2008})}\BibitemShut {NoStop}%
\bibitem [{\citenamefont {Rogers}\ \emph {et~al.}(2022)\citenamefont {Rogers},
  \citenamefont {Johnson},\ and\ \citenamefont {Munch}}]{rogers2022chaos}%
  \BibitemOpen
  \bibfield  {author} {\bibinfo {author} {\bibfnamefont {T.~L.}\ \bibnamefont
  {Rogers}}, \bibinfo {author} {\bibfnamefont {B.~J.}\ \bibnamefont
  {Johnson}},\ and\ \bibinfo {author} {\bibfnamefont {S.~B.}\ \bibnamefont
  {Munch}},\ }\href@noop {} {\bibfield  {journal} {\bibinfo  {journal} {Nature
  ecology \& evolution}\ }\textbf {\bibinfo {volume} {6}},\ \bibinfo {pages}
  {1105} (\bibinfo {year} {2022})}\BibitemShut {NoStop}%
\bibitem [{\citenamefont {Grilli}(2020)}]{grilli2020macroecological}%
  \BibitemOpen
  \bibfield  {author} {\bibinfo {author} {\bibfnamefont {J.}~\bibnamefont
  {Grilli}},\ }\href@noop {} {\bibfield  {journal} {\bibinfo  {journal} {Nature
  communications}\ }\textbf {\bibinfo {volume} {11}},\ \bibinfo {pages} {4743}
  (\bibinfo {year} {2020})}\BibitemShut {NoStop}%
\bibitem [{\citenamefont {Louca}\ \emph {et~al.}(2016)\citenamefont {Louca},
  \citenamefont {Jacques}, \citenamefont {Pires}, \citenamefont {Leal},
  \citenamefont {Srivastava}, \citenamefont {Parfrey}, \citenamefont
  {Farjalla},\ and\ \citenamefont {Doebeli}}]{louca2016high}%
  \BibitemOpen
  \bibfield  {author} {\bibinfo {author} {\bibfnamefont {S.}~\bibnamefont
  {Louca}}, \bibinfo {author} {\bibfnamefont {S.~M.}\ \bibnamefont {Jacques}},
  \bibinfo {author} {\bibfnamefont {A.~P.}\ \bibnamefont {Pires}}, \bibinfo
  {author} {\bibfnamefont {J.~S.}\ \bibnamefont {Leal}}, \bibinfo {author}
  {\bibfnamefont {D.~S.}\ \bibnamefont {Srivastava}}, \bibinfo {author}
  {\bibfnamefont {L.~W.}\ \bibnamefont {Parfrey}}, \bibinfo {author}
  {\bibfnamefont {V.~F.}\ \bibnamefont {Farjalla}},\ and\ \bibinfo {author}
  {\bibfnamefont {M.}~\bibnamefont {Doebeli}},\ }\href@noop {} {\bibfield
  {journal} {\bibinfo  {journal} {Nature ecology \& evolution}\ }\textbf
  {\bibinfo {volume} {1}},\ \bibinfo {pages} {0015} (\bibinfo {year}
  {2016})}\BibitemShut {NoStop}%
\bibitem [{\citenamefont {May}(1972)}]{may1972will}%
  \BibitemOpen
  \bibfield  {author} {\bibinfo {author} {\bibfnamefont {R.~M.}\ \bibnamefont
  {May}},\ }\href@noop {} {\bibfield  {journal} {\bibinfo  {journal} {Nature}\
  }\textbf {\bibinfo {volume} {238}},\ \bibinfo {pages} {413} (\bibinfo {year}
  {1972})}\BibitemShut {NoStop}%
\bibitem [{\citenamefont {Opper}\ and\ \citenamefont
  {Diederich}(1992)}]{opper1992phase}%
  \BibitemOpen
  \bibfield  {author} {\bibinfo {author} {\bibfnamefont {M.}~\bibnamefont
  {Opper}}\ and\ \bibinfo {author} {\bibfnamefont {S.}~\bibnamefont
  {Diederich}},\ }\href@noop {} {\bibfield  {journal} {\bibinfo  {journal}
  {Physical review letters}\ }\textbf {\bibinfo {volume} {69}},\ \bibinfo
  {pages} {1616} (\bibinfo {year} {1992})}\BibitemShut {NoStop}%
\bibitem [{\citenamefont {Bunin}(2017)}]{bunin2017ecological}%
  \BibitemOpen
  \bibfield  {author} {\bibinfo {author} {\bibfnamefont {G.}~\bibnamefont
  {Bunin}},\ }\href@noop {} {\bibfield  {journal} {\bibinfo  {journal}
  {Physical Review E}\ }\textbf {\bibinfo {volume} {95}},\ \bibinfo {pages}
  {042414} (\bibinfo {year} {2017})}\BibitemShut {NoStop}%
\bibitem [{\citenamefont {Blumenthal}\ \emph {et~al.}(2024)\citenamefont
  {Blumenthal}, \citenamefont {Rocks},\ and\ \citenamefont
  {Mehta}}]{blumenthal2024phase}%
  \BibitemOpen
  \bibfield  {author} {\bibinfo {author} {\bibfnamefont {E.}~\bibnamefont
  {Blumenthal}}, \bibinfo {author} {\bibfnamefont {J.~W.}\ \bibnamefont
  {Rocks}},\ and\ \bibinfo {author} {\bibfnamefont {P.}~\bibnamefont {Mehta}},\
  }\href@noop {} {\bibfield  {journal} {\bibinfo  {journal} {Physical review
  letters}\ }\textbf {\bibinfo {volume} {132}},\ \bibinfo {pages} {127401}
  (\bibinfo {year} {2024})}\BibitemShut {NoStop}%
\bibitem [{\citenamefont {Biroli}\ \emph {et~al.}(2018)\citenamefont {Biroli},
  \citenamefont {Bunin},\ and\ \citenamefont
  {Cammarota}}]{biroli2018marginally}%
  \BibitemOpen
  \bibfield  {author} {\bibinfo {author} {\bibfnamefont {G.}~\bibnamefont
  {Biroli}}, \bibinfo {author} {\bibfnamefont {G.}~\bibnamefont {Bunin}},\ and\
  \bibinfo {author} {\bibfnamefont {C.}~\bibnamefont {Cammarota}},\ }\href@noop
  {} {\bibfield  {journal} {\bibinfo  {journal} {New Journal of Physics}\
  }\textbf {\bibinfo {volume} {20}},\ \bibinfo {pages} {083051} (\bibinfo
  {year} {2018})}\BibitemShut {NoStop}%
\bibitem [{\citenamefont {Hu}\ \emph {et~al.}(2022)\citenamefont {Hu},
  \citenamefont {Amor}, \citenamefont {Barbier}, \citenamefont {Bunin},\ and\
  \citenamefont {Gore}}]{hu2022emergent}%
  \BibitemOpen
  \bibfield  {author} {\bibinfo {author} {\bibfnamefont {J.}~\bibnamefont
  {Hu}}, \bibinfo {author} {\bibfnamefont {D.~R.}\ \bibnamefont {Amor}},
  \bibinfo {author} {\bibfnamefont {M.}~\bibnamefont {Barbier}}, \bibinfo
  {author} {\bibfnamefont {G.}~\bibnamefont {Bunin}},\ and\ \bibinfo {author}
  {\bibfnamefont {J.}~\bibnamefont {Gore}},\ }\href@noop {} {\bibfield
  {journal} {\bibinfo  {journal} {Science}\ }\textbf {\bibinfo {volume}
  {378}},\ \bibinfo {pages} {85} (\bibinfo {year} {2022})}\BibitemShut
  {NoStop}%
\bibitem [{\citenamefont {Arnoulx~de Pirey}\ and\ \citenamefont
  {Bunin}(2024)}]{arnoulx2024many}%
  \BibitemOpen
  \bibfield  {author} {\bibinfo {author} {\bibfnamefont {T.}~\bibnamefont
  {Arnoulx~de Pirey}}\ and\ \bibinfo {author} {\bibfnamefont {G.}~\bibnamefont
  {Bunin}},\ }\href@noop {} {\bibfield  {journal} {\bibinfo  {journal}
  {Physical Review X}\ }\textbf {\bibinfo {volume} {14}},\ \bibinfo {pages}
  {011037} (\bibinfo {year} {2024})}\BibitemShut {NoStop}%
\bibitem [{\citenamefont {Arnoulx~de Pirey}\ and\ \citenamefont
  {Bunin}(2023)}]{arnoulx2023aging}%
  \BibitemOpen
  \bibfield  {author} {\bibinfo {author} {\bibfnamefont {T.}~\bibnamefont
  {Arnoulx~de Pirey}}\ and\ \bibinfo {author} {\bibfnamefont {G.}~\bibnamefont
  {Bunin}},\ }\href@noop {} {\bibfield  {journal} {\bibinfo  {journal}
  {Physical Review Letters}\ }\textbf {\bibinfo {volume} {130}},\ \bibinfo
  {pages} {098401} (\bibinfo {year} {2023})}\BibitemShut {NoStop}%
\bibitem [{\citenamefont {Pearce}\ \emph {et~al.}(2020)\citenamefont {Pearce},
  \citenamefont {Agarwala},\ and\ \citenamefont
  {Fisher}}]{pearce2020stabilization}%
  \BibitemOpen
  \bibfield  {author} {\bibinfo {author} {\bibfnamefont {M.~T.}\ \bibnamefont
  {Pearce}}, \bibinfo {author} {\bibfnamefont {A.}~\bibnamefont {Agarwala}},\
  and\ \bibinfo {author} {\bibfnamefont {D.~S.}\ \bibnamefont {Fisher}},\
  }\href@noop {} {\bibfield  {journal} {\bibinfo  {journal} {Proceedings of the
  National Academy of Sciences}\ }\textbf {\bibinfo {volume} {117}},\ \bibinfo
  {pages} {14572} (\bibinfo {year} {2020})}\BibitemShut {NoStop}%
\bibitem [{\citenamefont {Roy}\ \emph {et~al.}(2019)\citenamefont {Roy},
  \citenamefont {Biroli}, \citenamefont {Bunin},\ and\ \citenamefont
  {Cammarota}}]{roy2019numerical}%
  \BibitemOpen
  \bibfield  {author} {\bibinfo {author} {\bibfnamefont {F.}~\bibnamefont
  {Roy}}, \bibinfo {author} {\bibfnamefont {G.}~\bibnamefont {Biroli}},
  \bibinfo {author} {\bibfnamefont {G.}~\bibnamefont {Bunin}},\ and\ \bibinfo
  {author} {\bibfnamefont {C.}~\bibnamefont {Cammarota}},\ }\href@noop {}
  {\bibfield  {journal} {\bibinfo  {journal} {Journal of Physics A:
  Mathematical and Theoretical}\ }\textbf {\bibinfo {volume} {52}},\ \bibinfo
  {pages} {484001} (\bibinfo {year} {2019})}\BibitemShut {NoStop}%
\bibitem [{\citenamefont {Roy}\ \emph {et~al.}(2020)\citenamefont {Roy},
  \citenamefont {Barbier}, \citenamefont {Biroli},\ and\ \citenamefont
  {Bunin}}]{roy2020complex}%
  \BibitemOpen
  \bibfield  {author} {\bibinfo {author} {\bibfnamefont {F.}~\bibnamefont
  {Roy}}, \bibinfo {author} {\bibfnamefont {M.}~\bibnamefont {Barbier}},
  \bibinfo {author} {\bibfnamefont {G.}~\bibnamefont {Biroli}},\ and\ \bibinfo
  {author} {\bibfnamefont {G.}~\bibnamefont {Bunin}},\ }\href@noop {}
  {\bibfield  {journal} {\bibinfo  {journal} {PLoS computational biology}\
  }\textbf {\bibinfo {volume} {16}},\ \bibinfo {pages} {e1007827} (\bibinfo
  {year} {2020})}\BibitemShut {NoStop}%
\bibitem [{\citenamefont {Mallmin}\ \emph {et~al.}(2024)\citenamefont
  {Mallmin}, \citenamefont {Traulsen},\ and\ \citenamefont
  {De~Monte}}]{mallmin2024chaotic}%
  \BibitemOpen
  \bibfield  {author} {\bibinfo {author} {\bibfnamefont {E.}~\bibnamefont
  {Mallmin}}, \bibinfo {author} {\bibfnamefont {A.}~\bibnamefont {Traulsen}},\
  and\ \bibinfo {author} {\bibfnamefont {S.}~\bibnamefont {De~Monte}},\
  }\href@noop {} {\bibfield  {journal} {\bibinfo  {journal} {Proceedings of the
  National Academy of Sciences}\ }\textbf {\bibinfo {volume} {121}},\ \bibinfo
  {pages} {e2312822121} (\bibinfo {year} {2024})}\BibitemShut {NoStop}%
\bibitem [{\citenamefont {Leibold}\ \emph {et~al.}(2004)\citenamefont
  {Leibold}, \citenamefont {Holyoak}, \citenamefont {Mouquet}, \citenamefont
  {Amarasekare}, \citenamefont {Chase}, \citenamefont {Hoopes}, \citenamefont
  {Holt}, \citenamefont {Shurin}, \citenamefont {Law}, \citenamefont {Tilman}
  \emph {et~al.}}]{leibold2004metacommunity}%
  \BibitemOpen
  \bibfield  {author} {\bibinfo {author} {\bibfnamefont {M.~A.}\ \bibnamefont
  {Leibold}}, \bibinfo {author} {\bibfnamefont {M.}~\bibnamefont {Holyoak}},
  \bibinfo {author} {\bibfnamefont {N.}~\bibnamefont {Mouquet}}, \bibinfo
  {author} {\bibfnamefont {P.}~\bibnamefont {Amarasekare}}, \bibinfo {author}
  {\bibfnamefont {J.~M.}\ \bibnamefont {Chase}}, \bibinfo {author}
  {\bibfnamefont {M.~F.}\ \bibnamefont {Hoopes}}, \bibinfo {author}
  {\bibfnamefont {R.~D.}\ \bibnamefont {Holt}}, \bibinfo {author}
  {\bibfnamefont {J.~B.}\ \bibnamefont {Shurin}}, \bibinfo {author}
  {\bibfnamefont {R.}~\bibnamefont {Law}}, \bibinfo {author} {\bibfnamefont
  {D.}~\bibnamefont {Tilman}}, \emph {et~al.},\ }\href@noop {} {\bibfield
  {journal} {\bibinfo  {journal} {Ecology letters}\ }\textbf {\bibinfo {volume}
  {7}},\ \bibinfo {pages} {601} (\bibinfo {year} {2004})}\BibitemShut {NoStop}%
\bibitem [{\citenamefont {Garcia~Lorenzana}\ \emph {et~al.}(2024)\citenamefont
  {Garcia~Lorenzana}, \citenamefont {Altieri},\ and\ \citenamefont
  {Biroli}}]{garcia2024interactions}%
  \BibitemOpen
  \bibfield  {author} {\bibinfo {author} {\bibfnamefont {G.}~\bibnamefont
  {Garcia~Lorenzana}}, \bibinfo {author} {\bibfnamefont {A.}~\bibnamefont
  {Altieri}},\ and\ \bibinfo {author} {\bibfnamefont {G.}~\bibnamefont
  {Biroli}},\ }\href@noop {} {\bibfield  {journal} {\bibinfo  {journal} {PRX
  Life}\ }\textbf {\bibinfo {volume} {2}},\ \bibinfo {pages} {013014} (\bibinfo
  {year} {2024})}\BibitemShut {NoStop}%
\bibitem [{\citenamefont {Loreau}\ \emph {et~al.}(2003)\citenamefont {Loreau},
  \citenamefont {Mouquet},\ and\ \citenamefont
  {Gonzalez}}]{loreau2003biodiversity}%
  \BibitemOpen
  \bibfield  {author} {\bibinfo {author} {\bibfnamefont {M.}~\bibnamefont
  {Loreau}}, \bibinfo {author} {\bibfnamefont {N.}~\bibnamefont {Mouquet}},\
  and\ \bibinfo {author} {\bibfnamefont {A.}~\bibnamefont {Gonzalez}},\
  }\href@noop {} {\bibfield  {journal} {\bibinfo  {journal} {Proceedings of the
  National Academy of Sciences}\ }\textbf {\bibinfo {volume} {100}},\ \bibinfo
  {pages} {12765} (\bibinfo {year} {2003})}\BibitemShut {NoStop}%
\bibitem [{\citenamefont {Hinrichsen}(2000)}]{hinrichsen2000non}%
  \BibitemOpen
  \bibfield  {author} {\bibinfo {author} {\bibfnamefont {H.}~\bibnamefont
  {Hinrichsen}},\ }\href@noop {} {\bibfield  {journal} {\bibinfo  {journal}
  {Advances in physics}\ }\textbf {\bibinfo {volume} {49}},\ \bibinfo {pages}
  {815} (\bibinfo {year} {2000})}\BibitemShut {NoStop}%
\bibitem [{\citenamefont {Krapivsky}\ \emph {et~al.}(2010)\citenamefont
  {Krapivsky}, \citenamefont {Redner},\ and\ \citenamefont
  {Ben-Naim}}]{krapivsky2010kinetic}%
  \BibitemOpen
  \bibfield  {author} {\bibinfo {author} {\bibfnamefont {P.~L.}\ \bibnamefont
  {Krapivsky}}, \bibinfo {author} {\bibfnamefont {S.}~\bibnamefont {Redner}},\
  and\ \bibinfo {author} {\bibfnamefont {E.}~\bibnamefont {Ben-Naim}},\
  }\href@noop {} {\emph {\bibinfo {title} {A kinetic view of statistical
  physics}}}\ (\bibinfo  {publisher} {Cambridge University Press},\ \bibinfo
  {year} {2010})\BibitemShut {NoStop}%
\bibitem [{\citenamefont {Sompolinsky}\ and\ \citenamefont
  {Zippelius}(1982)}]{sompolinsky1982relaxational}%
  \BibitemOpen
  \bibfield  {author} {\bibinfo {author} {\bibfnamefont {H.}~\bibnamefont
  {Sompolinsky}}\ and\ \bibinfo {author} {\bibfnamefont {A.}~\bibnamefont
  {Zippelius}},\ }\href@noop {} {\bibfield  {journal} {\bibinfo  {journal}
  {Physical Review B}\ }\textbf {\bibinfo {volume} {25}},\ \bibinfo {pages}
  {6860} (\bibinfo {year} {1982})}\BibitemShut {NoStop}%
\bibitem [{\citenamefont {Galla}(2024)}]{galla2024generating}%
  \BibitemOpen
  \bibfield  {author} {\bibinfo {author} {\bibfnamefont {T.}~\bibnamefont
  {Galla}},\ }\href@noop {} {\bibfield  {journal} {\bibinfo  {journal} {arXiv
  preprint arXiv:2405.14289}\ } (\bibinfo {year} {2024})}\BibitemShut {NoStop}%
\bibitem [{\citenamefont {Grinstein}\ and\ \citenamefont
  {Mu{\~n}oz}(1997)}]{grinstein1997statistical}%
  \BibitemOpen
  \bibfield  {author} {\bibinfo {author} {\bibfnamefont {G.}~\bibnamefont
  {Grinstein}}\ and\ \bibinfo {author} {\bibfnamefont {M.~A.}\ \bibnamefont
  {Mu{\~n}oz}},\ }in\ \href@noop {} {\emph {\bibinfo {booktitle} {Fourth
  Granada Lectures in Computational Physics}}}\ (\bibinfo  {publisher}
  {Springer},\ \bibinfo {year} {1997})\ pp.\ \bibinfo {pages}
  {223--270}\BibitemShut {NoStop}%
\bibitem [{\citenamefont {Ottino-L{\"o}ffler}\ and\ \citenamefont
  {Kardar}(2020)}]{ottino2020population}%
  \BibitemOpen
  \bibfield  {author} {\bibinfo {author} {\bibfnamefont {B.}~\bibnamefont
  {Ottino-L{\"o}ffler}}\ and\ \bibinfo {author} {\bibfnamefont
  {M.}~\bibnamefont {Kardar}},\ }\href@noop {} {\bibfield  {journal} {\bibinfo
  {journal} {Physical Review E}\ }\textbf {\bibinfo {volume} {102}},\ \bibinfo
  {pages} {052106} (\bibinfo {year} {2020})}\BibitemShut {NoStop}%
\bibitem [{\citenamefont {Van~Wijland}\ \emph {et~al.}(1998)\citenamefont
  {Van~Wijland}, \citenamefont {Oerding},\ and\ \citenamefont
  {Hilhorst}}]{van1998wilson}%
  \BibitemOpen
  \bibfield  {author} {\bibinfo {author} {\bibfnamefont {F.}~\bibnamefont
  {Van~Wijland}}, \bibinfo {author} {\bibfnamefont {K.}~\bibnamefont
  {Oerding}},\ and\ \bibinfo {author} {\bibfnamefont {H.}~\bibnamefont
  {Hilhorst}},\ }\href@noop {} {\bibfield  {journal} {\bibinfo  {journal}
  {Physica A: Statistical Mechanics and its Applications}\ }\textbf {\bibinfo
  {volume} {251}},\ \bibinfo {pages} {179} (\bibinfo {year}
  {1998})}\BibitemShut {NoStop}%
\bibitem [{\citenamefont {Kardar}\ \emph {et~al.}(1986)\citenamefont {Kardar},
  \citenamefont {Parisi},\ and\ \citenamefont {Zhang}}]{kardar1986dynamic}%
  \BibitemOpen
  \bibfield  {author} {\bibinfo {author} {\bibfnamefont {M.}~\bibnamefont
  {Kardar}}, \bibinfo {author} {\bibfnamefont {G.}~\bibnamefont {Parisi}},\
  and\ \bibinfo {author} {\bibfnamefont {Y.-C.}\ \bibnamefont {Zhang}},\
  }\href@noop {} {\bibfield  {journal} {\bibinfo  {journal} {Physical Review
  Letters}\ }\textbf {\bibinfo {volume} {56}},\ \bibinfo {pages} {889}
  (\bibinfo {year} {1986})}\BibitemShut {NoStop}%
\bibitem [{\citenamefont {Arnoulx~de Pirey}\ and\ \citenamefont
  {Bunin}(2025)}]{arnoulx2025critical}%
  \BibitemOpen
  \bibfield  {author} {\bibinfo {author} {\bibfnamefont {T.}~\bibnamefont
  {Arnoulx~de Pirey}}\ and\ \bibinfo {author} {\bibfnamefont {G.}~\bibnamefont
  {Bunin}},\ }\href@noop {} {\bibfield  {journal} {\bibinfo  {journal} {SciPost
  Physics}\ }\textbf {\bibinfo {volume} {18}},\ \bibinfo {pages} {051}
  (\bibinfo {year} {2025})}\BibitemShut {NoStop}%
\bibitem [{\citenamefont {Medina}\ \emph {et~al.}(1989)\citenamefont {Medina},
  \citenamefont {Hwa}, \citenamefont {Kardar},\ and\ \citenamefont
  {Zhang}}]{medina1989burgers}%
  \BibitemOpen
  \bibfield  {author} {\bibinfo {author} {\bibfnamefont {E.}~\bibnamefont
  {Medina}}, \bibinfo {author} {\bibfnamefont {T.}~\bibnamefont {Hwa}},
  \bibinfo {author} {\bibfnamefont {M.}~\bibnamefont {Kardar}},\ and\ \bibinfo
  {author} {\bibfnamefont {Y.-C.}\ \bibnamefont {Zhang}},\ }\href@noop {}
  {\bibfield  {journal} {\bibinfo  {journal} {Physical Review A}\ }\textbf
  {\bibinfo {volume} {39}},\ \bibinfo {pages} {3053} (\bibinfo {year}
  {1989})}\BibitemShut {NoStop}%
\bibitem [{\citenamefont {Janssen}\ \emph {et~al.}(1999)\citenamefont
  {Janssen}, \citenamefont {T{\"a}uber},\ and\ \citenamefont
  {Frey}}]{janssen1999exact}%
  \BibitemOpen
  \bibfield  {author} {\bibinfo {author} {\bibfnamefont {H.}~\bibnamefont
  {Janssen}}, \bibinfo {author} {\bibfnamefont {U.~C.}\ \bibnamefont
  {T{\"a}uber}},\ and\ \bibinfo {author} {\bibfnamefont {E.}~\bibnamefont
  {Frey}},\ }\href@noop {} {\bibfield  {journal} {\bibinfo  {journal} {The
  European Physical Journal B-Condensed Matter and Complex Systems}\ }\textbf
  {\bibinfo {volume} {9}},\ \bibinfo {pages} {491} (\bibinfo {year}
  {1999})}\BibitemShut {NoStop}%
\bibitem [{sup()}]{supp}%
  \BibitemOpen
  \href@noop {} {}\bibinfo {note} {See Supplemental Material for mathematical
  details, which includes
  Refs.~\cite{janssen1976lagrangean,feller1951two,dornic2005integration,weissmann2018simulation}.}\BibitemShut
  {Stop}%
\bibitem [{\citenamefont {Lynch}\ and\ \citenamefont
  {Neufeld}(2015)}]{lynch2015ecology}%
  \BibitemOpen
  \bibfield  {author} {\bibinfo {author} {\bibfnamefont {M.~D.}\ \bibnamefont
  {Lynch}}\ and\ \bibinfo {author} {\bibfnamefont {J.~D.}\ \bibnamefont
  {Neufeld}},\ }\href@noop {} {\bibfield  {journal} {\bibinfo  {journal}
  {Nature Reviews Microbiology}\ }\textbf {\bibinfo {volume} {13}},\ \bibinfo
  {pages} {217} (\bibinfo {year} {2015})}\BibitemShut {NoStop}%
\bibitem [{\citenamefont {Denk}\ and\ \citenamefont
  {Hallatschek}(2022)}]{denk2022self}%
  \BibitemOpen
  \bibfield  {author} {\bibinfo {author} {\bibfnamefont {J.}~\bibnamefont
  {Denk}}\ and\ \bibinfo {author} {\bibfnamefont {O.}~\bibnamefont
  {Hallatschek}},\ }\href@noop {} {\bibfield  {journal} {\bibinfo  {journal}
  {Proceedings of the National Academy of Sciences}\ }\textbf {\bibinfo
  {volume} {119}},\ \bibinfo {pages} {e2200390119} (\bibinfo {year}
  {2022})}\BibitemShut {NoStop}%
\bibitem [{\citenamefont {Janssen}(1997)}]{janssen1997renormalized}%
  \BibitemOpen
  \bibfield  {author} {\bibinfo {author} {\bibfnamefont {H.}~\bibnamefont
  {Janssen}},\ }\href@noop {} {\bibfield  {journal} {\bibinfo  {journal}
  {Physical Review E}\ }\textbf {\bibinfo {volume} {55}},\ \bibinfo {pages}
  {6253} (\bibinfo {year} {1997})}\BibitemShut {NoStop}%
\bibitem [{\citenamefont {Hooyberghs}\ \emph {et~al.}(2004)\citenamefont
  {Hooyberghs}, \citenamefont {Igl{\'o}i},\ and\ \citenamefont
  {Vanderzande}}]{hooyberghs2004absorbing}%
  \BibitemOpen
  \bibfield  {author} {\bibinfo {author} {\bibfnamefont {J.}~\bibnamefont
  {Hooyberghs}}, \bibinfo {author} {\bibfnamefont {F.}~\bibnamefont
  {Igl{\'o}i}},\ and\ \bibinfo {author} {\bibfnamefont {C.}~\bibnamefont
  {Vanderzande}},\ }\href@noop {} {\bibfield  {journal} {\bibinfo  {journal}
  {Physical Review E -- Statistical, Nonlinear, and Soft Matter Physics}\
  }\textbf {\bibinfo {volume} {69}},\ \bibinfo {pages} {066140} (\bibinfo
  {year} {2004})}\BibitemShut {NoStop}%
\bibitem [{\citenamefont {Noest}(1986)}]{noest1986new}%
  \BibitemOpen
  \bibfield  {author} {\bibinfo {author} {\bibfnamefont {A.~J.}\ \bibnamefont
  {Noest}},\ }\href@noop {} {\bibfield  {journal} {\bibinfo  {journal}
  {Physical review letters}\ }\textbf {\bibinfo {volume} {57}},\ \bibinfo
  {pages} {90} (\bibinfo {year} {1986})}\BibitemShut {NoStop}%
\bibitem [{\citenamefont {Janssen}(1976)}]{janssen1976lagrangean}%
  \BibitemOpen
  \bibfield  {author} {\bibinfo {author} {\bibfnamefont {H.-K.}\ \bibnamefont
  {Janssen}},\ }\href@noop {} {\bibfield  {journal} {\bibinfo  {journal}
  {Zeitschrift f{\"u}r Physik B Condensed Matter}\ }\textbf {\bibinfo {volume}
  {23}},\ \bibinfo {pages} {377} (\bibinfo {year} {1976})}\BibitemShut
  {NoStop}%
\bibitem [{\citenamefont {Feller}(1951)}]{feller1951two}%
  \BibitemOpen
  \bibfield  {author} {\bibinfo {author} {\bibfnamefont {W.}~\bibnamefont
  {Feller}},\ }\href@noop {} {\bibfield  {journal} {\bibinfo  {journal} {Annals
  of mathematics}\ }\textbf {\bibinfo {volume} {54}},\ \bibinfo {pages} {173}
  (\bibinfo {year} {1951})}\BibitemShut {NoStop}%
\bibitem [{\citenamefont {Dornic}\ \emph {et~al.}(2005)\citenamefont {Dornic},
  \citenamefont {Chat{\'e}},\ and\ \citenamefont
  {Munoz}}]{dornic2005integration}%
  \BibitemOpen
  \bibfield  {author} {\bibinfo {author} {\bibfnamefont {I.}~\bibnamefont
  {Dornic}}, \bibinfo {author} {\bibfnamefont {H.}~\bibnamefont {Chat{\'e}}},\
  and\ \bibinfo {author} {\bibfnamefont {M.~A.}\ \bibnamefont {Munoz}},\
  }\href@noop {} {\bibfield  {journal} {\bibinfo  {journal} {Physical review
  letters}\ }\textbf {\bibinfo {volume} {94}},\ \bibinfo {pages} {100601}
  (\bibinfo {year} {2005})}\BibitemShut {NoStop}%
\bibitem [{\citenamefont {Weissmann}\ \emph {et~al.}(2018)\citenamefont
  {Weissmann}, \citenamefont {Shnerb},\ and\ \citenamefont
  {Kessler}}]{weissmann2018simulation}%
  \BibitemOpen
  \bibfield  {author} {\bibinfo {author} {\bibfnamefont {H.}~\bibnamefont
  {Weissmann}}, \bibinfo {author} {\bibfnamefont {N.~M.}\ \bibnamefont
  {Shnerb}},\ and\ \bibinfo {author} {\bibfnamefont {D.~A.}\ \bibnamefont
  {Kessler}},\ }\href@noop {} {\bibfield  {journal} {\bibinfo  {journal}
  {Physical Review E}\ }\textbf {\bibinfo {volume} {98}},\ \bibinfo {pages}
  {022131} (\bibinfo {year} {2018})}\BibitemShut {NoStop}%
\end{thebibliography}%

\clearpage
\onecolumngrid       
\section*{Supplemental material for ``Self-organized criticality in complex
model ecosystems''}

\subsection*{Stability of directed percolation scaling under the self-consistency
condition}

\noindent Here we consider the directed percolation field theory in
the presence of long-range environmental noise, 
\begin{equation}
\partial_{t}n=n(\bar{g}+\delta g(x,t)-n)+D\Delta n+\sqrt{2Tn}\,\eta(x,t)\,.\label{eq:field_theory_SM-1}
\end{equation}

\noindent We denote $C_{0}(x,t)=\left\langle \delta g(x,t)\delta g(0,0)\right\rangle $
and assume the large-scale form
\begin{equation}
C_{0}(x,t)\sim A_{0}x^{-2\mu}\mathcal{G}\left(tx^{-z}\right)\,,\label{eq:correlations}
\end{equation}

\noindent where $z$ is the dynamic critical exponent of directed
percolation in the corresponding dimension. The exponent $\mu$ is such that $2\mu<d+z$, making the correlation function
in Eq.~(\ref{eq:correlations}) non-integrable. We use the Martin-Siggia-Rose-Janssen-de
Dominicis formalism \citep{janssen1976lagrangean}, and write the
generating functional in terms of the original field $n$ and a conjugated
response field $\hat{n}$ with an action given by

\begin{align}
\mathcal{S}_{0}[n,\hat{n}] & =\int dx\,dt\,\left[\hat{n}\left(\partial_{t}n-n\bar{g}-D\Delta n\right)+\sqrt{T}\hat{n}n\left(n-\hat{n}\right)\right]\label{eq:action_MSRJD}\\
 & -\frac{1}{2}\int dx\,dt\int dx'\,dt'\,n(x,t)n(x',t')\hat{n}(x,t)\hat{n}(x',t')C_{0}(x-x',t-t')\,.\nonumber 
\end{align}

\noindent This action (where the Ito convention was used) was obtained
after the rescaling $n\to\sqrt{T}n$, which makes the time-reversal
symmetry $n(x,t)\to-\hat{n}(x,-t)$ and $\hat{n}(x,t)\to-n(x,-t)$
explicit. The last term originates from the environmental noise $\delta g(x,t)$.
At the directed percolation fixed point, the theory is left invariant
upon the integration of high-momentum spatial Fourier modes belonging
to the shell $\Lambda{\rm e}^{-\delta\ell}<|k|<\Lambda$ followed
by the rescaling $x\to{\rm e}^{\delta\ell}x'$, $t\to{\rm e}^{z\delta\ell}t'$,
$n\to{\rm e}^{-\zeta\delta\ell}n'$ and $\hat{n}\to{\rm e}^{-\zeta\delta\ell}\hat{n}'$
for $\delta\ell\ll1$. Under this rescaling alone, the amplitude of
the large-scale behavior of the correlation function $C_{0}(x,t)$
is changed according to
\[
A_{0}\to A_{\delta\ell}=A_{0}{\rm e}^{\delta\ell(2d+2z-4\zeta-2\mu)}\,.
\]

\noindent Therefore, under the directed percolation scaling, the renormalization
flow of the large-scale amplitude of the environmental noise is of
the form
\begin{equation}
\partial_{\ell}A_{\ell}=(2d+2z-4\zeta-2\mu)A_{\ell}+\text{graphical corrections}\,,\label{eq:RG_flow}
\end{equation}

\noindent where the graphical corrections account for the integration
over the high-momentum shell. As we show here, there is no first order
(in $A_{\ell}$) graphical corrections to the flow equation of $A_{\ell}$.
This guarantees the stability of the directed percolation scaling
in the presence of weak long-range environmental noise provided that
\[
\mu>z+d-2\zeta\,,
\]

\noindent as claimed in the main text. On the other hand, if $\mu<z+d-2\zeta$,
any amount of long-range environmental noise is expected to change
the scaling behavior. In the case of the Kardar-Parisi-Zhang (KPZ)
field theory \citep{kardar1986dynamic} driven by a long-range spatially
correlated noise, this argument was used in \citep{janssen1999exact}
to derive a condition for the stability of the KPZ universality
class. When the noise driving the KPZ equation exhibits long-range
correlations in both space and time, the same argument applies and
we have checked that it allows to recover the bound originally
derived in \citep{medina1989burgers}. 

\medskip{}

\begin{figure}[h!]
\begin{center} 
\begin{overpic}[totalheight=2cm]{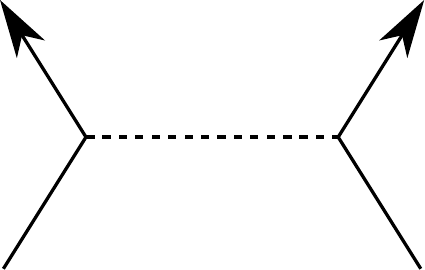}
\put (-5,30) {$x,\,t$}  
\put (85,30) {$x' ,\, t'$}
 
\end{overpic}
\caption{Vertex associated with the environmental noise $\delta g(x,t)$. An outgoing leg with an arrow corresponds to a response field $\hat{n}$ and an outgoing leg with no arrow corresponds to a field $n$.}
\label{fig:feynman-1}
\end{center}
\end{figure}

\begin{figure}[h!]
\begin{center} 
\begin{overpic}[totalheight=3cm]{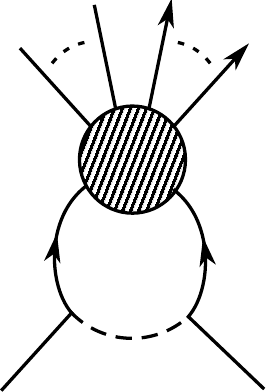}
\put (-15,-10) {$k_1,\,\omega_1$}  
\put (55,-10) {$k_2 ,\, \omega_2$}
\put (-20,92) {$k_3 ,\, \omega_3$}
\put (-5,105) {$k_p ,\, \omega_p$}
\put (45,105) {$\tilde{k}_1 ,\, \tilde{\omega}_1$}
\put (60,92) {$\tilde{k}_q ,\, \tilde{\omega}_q$}
 
\end{overpic}
\caption{Fourier-space representation of the contraction of the vertex arising from the environmental noise and the vertex with $(p,q)$ externel legs of the directed percolation theory at the renormalization scale $\ell$. A continuous line with an arrow denotes the Gaussian propagator $G_{\ell}(k,\omega)$ of the directed percolation theory  at the same renormalization scale.}
\label{fig:feynman-2}
\end{center}
\end{figure}
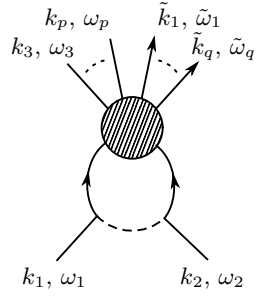

\noindent To proceed with the investigation of the graphical corrections,
we introduce the vertex associated to the second line of Eq.~(\ref{eq:action_MSRJD})
in Fig.~\ref{fig:feynman-1}. To first order, the graphical corrections
are made from the contraction of one vertex shown in Fig.~\ref{fig:feynman-1}
with any subset of the vertices generated by the renormalization flow
of the original directed percolation action given by the first line
of Eq.~(\ref{eq:action_MSRJD}). The resulting diagrams must be one-particle
irreducible and have a single loop (because the internal lines all
carry momenta belonging to the infinitesimal high-momentum shell).
First, if the dotted line belongs to the loop, the resulting graph shows no small-momenta divergence. To illustrate this, consider the diagram in Fig.~\ref{fig:feynman-2},
which is a contraction of the vertex in Fig.~\ref{fig:feynman-1}
with the directed percolation vertex that possesses $(p,q)$ external
legs, where $p\geq2$ is the number of external $n$ fields
and $q\geq0$ the number of external $\hat{n}$ fields. We denote the Fourier amplitude of this
vertex by 
\[
I_{(p,q)}^{\ell}\left(\{k_{i},\omega_{i}\}_{i=1\dots p};\{\hat{k}_{j},\hat{\omega}_{j}\}_{j=1\dots q}\right)\,,
\]
where $\{k_{i},\omega_{i}\}_{i=1\dots p}$ are the momenta of the
$n$ fields and $\{\hat{k}_{j},\hat{\omega}_{j}\}_{j=1\dots m}$ that
of the $\hat{n}$ fields.

\noindent 

\begin{figure}[h!]
\begin{center} 
\begin{overpic}[totalheight=2cm]{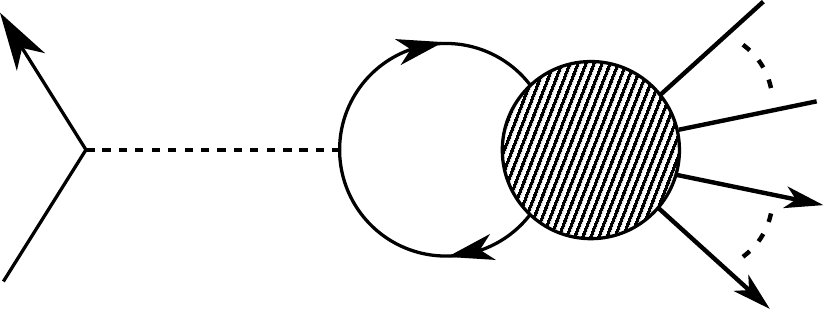}
 
\end{overpic}
\caption{A typical 1-particle irreducible graph where the dotted line is not part of the loop. One can go around the loop in the direction of the arrow, so the graph vanishes by causality.}
\label{fig:feynman-3}
\end{center}
\end{figure}

\begin{figure}[h!]
\begin{center} 
\begin{overpic}[totalheight=2cm]{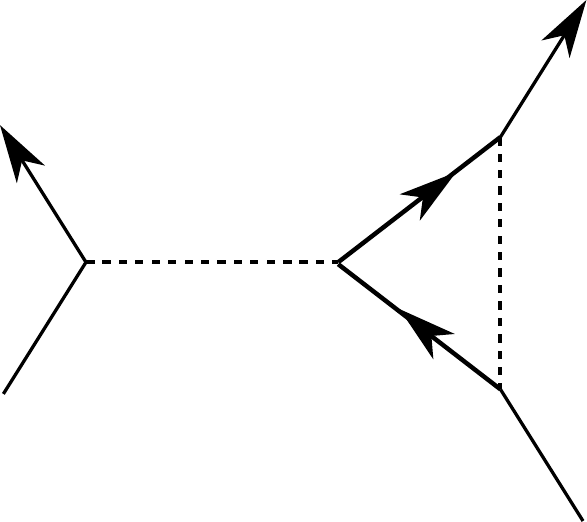}
 
\end{overpic}
\caption{A second-order graph which introduces graphical corrections to the flow equation of the large-scale amplitude $A_\ell$.}
\label{fig:feynman-4}
\end{center}
\end{figure}

The diagram in Fig.~\ref{fig:feynman-2} enters the first-order in $A_{\ell}$ renormalization of the amplitude
$I_{(p,q)}^{\ell}$ through a graphical correction that we denote
$\Gamma_{(p,q)}^{\ell}$. As a function of the external momenta shown
in Fig.~\ref{fig:feynman-2}, its expression is given by
\begin{align*}
\Gamma_{(p,q)}^{\ell}\left(\{k_{i},\omega_{i}\}_{i=1\dots n};\{\hat{k}_{j},\hat{\omega}_{j}\}_{j=1\dots m}\right) & =\binom{p}{2}\frac{1}{\delta\ell}\int_{d\Omega}dk\int d\omega\,G^{\ell}(-k,-\omega)G^{\ell}(k-(k_{1}+k_{2}),\omega-(\omega_{1}+\omega_{2}))\hat{C}_{\ell}(k+k_{1},\omega+\omega_{1})\times\\
 & \dots\times I_{(p,q)}^{\ell}\left(k,k_{1}+k_{2}-k,\{k_{i},\omega_{i}\}_{i=3\dots p};\{\hat{k}_{j},\hat{\omega}_{j}\}_{j=1\dots q}\right)
\end{align*}
where $G^{\ell}(k,\omega)$ is the Gaussian propagator of the directed
percolation action at renormalization scale $\ell$ and where $d\Omega$
denotes the high-momentum infinitesimal shell $\Lambda{\rm e}^{-\delta\ell}<|k|<\Lambda$.
Crucially, this graphical correction is well-behaved as the external
momenta are set to zero. In fact,
\[
\Gamma_{(p,q)}^{\ell}\left(\{0,0\}_{i=1\dots n};\{0,0\}_{j=1\dots m}\right)=\binom{p}{2}\frac{1}{\delta\ell}\int_{d\Omega}dk\int d\omega\,G^{\ell}(-k,-\omega)G^{\ell}(k,\omega)\hat{C}_{\ell}(k,\omega)I_{(p,q)}^{\ell}\left(k,-k,\{0,0\}_{i=3\dots p};\{0,0\}_{j=1\dots q}\right)\,,
\]

\noindent is finite since the $k$-integral restricts the evaluation
of $\hat{C}_{\ell}(k,\omega)$ to a high-momenta region where it is
non-singular, and since we expect the directed percolation vertices to be well-behaved at small momenta and large $\ell$. Therefore,
to first order in $A_{\ell}$, a graphical correction that diverges at small momenta must necessarily arise from a graph where
the dotted line is not part of the loop. Such graphs, however, vanish
by causality, see Fig.~\ref{fig:feynman-3}. Therefore, there are
no graphical corrections to the renormalization flow in Eq.~(\ref{eq:RG_flow}),
as claimed above. Note that this breaks down to second order in $A_{\ell}$,
see for instance the graph in Fig.~\ref{fig:feynman-4}. 

\noindent 

\noindent 

\subsection*{Scaling behavior above $d_{c}=4$}

\noindent We consider the directed percolation field theory, at $d>4$,
in the presence of short-range (here taken white) environmental noise,
\begin{equation}
\partial_{t}n=n(\bar{g}+\delta g(x,t)-n)+D\Delta n+\sqrt{2Tn}\,\eta(x,t)\,.\label{eq:field_theory_SM}
\end{equation}

\noindent We proceed with the rescaling $x\to bx'$, $t\to b^{2}t'$
and $n\to b^{-2}n'$ which leaves us with
\begin{equation}
\partial_{t'}n'=n'\left(\bar{g}b^{2}+b^{1-d/2}\delta g'(x',t')-n'\right)+D\Delta'n'+b^{2-d/2}\sqrt{2Tn'}\,\eta'(x',t')\,,\label{eq:rescaled_EOM}
\end{equation}

\noindent where $\delta g'$ and $\eta'$ are unit variance Gaussian
white noises of the new coordinates. We take $b=\bar{g}^{-1/2}$ in
the following. To leading order as $\bar{g}\to0$, Eq.~(\ref{eq:rescaled_EOM})
becomes noiseless
\[
\partial_{t'}n'=n'\left(1-n'\right)+D\Delta'n'\,,
\]

\noindent showing that 
\[
\left\langle n'\right\rangle \to1\,,
\]

\noindent in that limit. To next-to-leading order, we get the expansion
$n'=1+\bar{g}^{d/4-1}u$ where 
\[
\partial_{t'}u=-u+D\Delta'u+\sqrt{2T}\,\eta'(x',t')\,.
\]

\noindent Therefore, we obtain
\[
\left\langle u(x',t')u(0,0)\right\rangle =C_{0}(x',t')
\]

\noindent with in Fourier space
\[
\hat{C}_{0}(k,\omega)=\frac{2T}{\omega^{2}+\left(Dk^{2}+1\right)^{2}}\,.
\]

\noindent Hence, close to the critical point $\bar{g}\to0$, we get
the scaling form
\[
\left\langle \delta n(x,t)\delta n(0,0)\right\rangle _{\bar{g}}\sim\bar{g}^{d/2}C_{0}\left(\bar{g}^{1/2}x,\bar{g}t\right)\,.
\]

\noindent Integrating over $\bar{g}$ then yields Eq.~(6) of the
main text. 

\subsection*{Large-scale behavior on the fully connected lattice}

\noindent We consider here the extension of our work to the fully
connected lattice with $M\gg1$ sites (labeled by Greek indices $\mu,\,\nu$).
We thus study the set of (Ito) equations 
\begin{equation}
\dot{n}_{\mu}=n_{\mu}(\bar{g}+\delta g_{\mu}(t)-n_{\mu})+D(m(t)-n_{\mu})+\sqrt{2n_{\mu}}\eta_{\mu}(t)\,,\label{eq:DP_fully_connected_many_body}
\end{equation}

\noindent where 
\[
m(t)=\frac{1}{M}\sum_{\nu}n_{\nu}(t)\,,
\]

\noindent ensures an all-to-all coupling between the sites. The $\eta_{\mu}(t)$
are uncorrelated Gaussian white noises $\left\langle \eta_{\mu}(t)\eta_{\nu}(t')\right\rangle =\delta_{\mu\nu}\delta(t-t')$.
Furthermore, the environmental noises $\delta g_{\mu}(t)$ are a set
of Gaussian processes with correlations given by the solution of the
self-consistency condition
\begin{equation}
\left\langle \delta g_{\mu}(t)\delta g_{\nu}(t')\right\rangle =\sigma^{2}\int d\bar{g}\,P(\bar{g})\left\langle \left(n_{\mu}(t)-\left\langle m\right\rangle _{\bar{g}}\right)\left(n_{\nu}(t')-\left\langle m\right\rangle _{\bar{g}}\right)\right\rangle _{\bar{g}}\,,\label{eq:self_consistency_fully_connected}
\end{equation}
with $P(\bar{g})$ a Gaussian distribution with mean and (strictly
positive) variance also determined self-consistently, see Eq.~(3)
of the main text. The solution to Eq.~(\ref{eq:self_consistency_fully_connected}),
which only depends on whether $\mu=\nu$ or $\mu\neq\nu$, can be
viewed as the fixed point of an iterative process: Start with an ansatz
for the correlations of the environmental noise, derive the corresponding
fluctuations of populations sizes, and use them to update the correlations
of the noise. Here, we take the simplest initial ansatz $\delta g_{\mu}(t)=0$.
As in the finite dimensional case, we show that the resulting two-time
correlations of population sizes exhibit power-law tails after integrating
over the average growth rate $\bar{g}$. However, their amplitude
decays with the system size as $M^{-1}$. 

We start by focusing on the statistics of $m(t)$ at a given value
of $\bar{g}$. We assume furthermore that the initial conditions $n_{\mu}(0)$
are independently distributed from a distribution $p(x)$ with support
on $x>0$. For any given observable $\mathcal{O}[m]$, we have the
large-deviation form
\begin{align}
\left\langle \mathcal{O}[m]\right\rangle  & =\int\mathcal{D}\hat{m}\,\mathcal{O}[m]\exp\left(-M\mathcal{S}[\hat{m},m]\right)\label{eq:action_DP_fully_connected}
\end{align}

\noindent with the action
\[
\mathcal{S}[\hat{m},m]=-i\int_{0}^{t}\hat{m}(s)m(s)-\ln\left\langle \exp\left(-i\int_{0}^{t}\hat{m}(s)n(s)\right)\right\rangle \,,
\]

\noindent and where the average $\left\langle \dots\right\rangle $
refers to an average over the stochastic process
\[
\dot{n}=n(\bar{g}-n)+D(m(s)-n)+\sqrt{2n}\eta(t)\,,
\]

\noindent with initial condition $n(0)=x$ drawn from the distribution
$p(x)$. To leading order as $M\to\infty$, the evolution of $m(s)$
is deterministic and is given by the saddle-point equation
\begin{align*}
\frac{\delta\mathcal{S}}{\delta\hat{m}(s)} & =0\,,\\
\frac{\delta\mathcal{S}}{\delta m(s)} & =0\,.
\end{align*}

\noindent We have
\[
\frac{\delta\mathcal{S}}{\delta m(s)}=-i\hat{m}(s)-\frac{\delta}{\delta m(s)}\ln\left\langle \exp\left(-i\int_{0}^{t}\hat{m}(s)n(s)\right)\right\rangle \,,
\]

\noindent and get, as expected for the deterministic trajectory, that
\[
\frac{\delta\mathcal{S}}{\delta m(s)}=0\,\,\,\text{for}\,\,\,\hat{m}=0\,.
\]

\noindent We are thus left with the equation
\[
0=\frac{\delta\mathcal{S}}{\delta\hat{m}(s)}=-im(s)+i\frac{\left\langle n(s)\exp\left(-i\int_{0}^{t}\hat{m}(s')n(s')\right)\right\rangle }{\left\langle \exp\left(-i\int_{0}^{t}\hat{m}(s')n(s')\right)\right\rangle }\,,
\]

\noindent which, for $\hat{m}=0$, reduces to
\begin{equation}
m(s)=\left\langle n(s)\right\rangle \,.\label{eq:DP_full_self_consistency}
\end{equation}

\noindent In the steady state, the saddle-point solution converges to the mean value $\bar{m}$. The static properties
in the steady state have been studied in \citep{ottino2020population},
and we quote some of their results in the following. 

\subsubsection*{The steady state of the saddle-point equation}

\noindent In the steady state, we look for a solution of the equation
$\bar{m}=\langle n\rangle$, where the dynamics of $n(t)$ is given
by the following equation
\begin{equation}
\dot{n}=n(\bar{g}-n)+\sqrt{2n}\eta(t)+D(\bar{m}-n)\,.\label{eq:DP_fully_connected}
\end{equation}

\noindent The steady-state distribution associated to Eq.~(\ref{eq:DP_fully_connected})
is given by
\[
P_{\bar{g}}(n)=\frac{1}{Z}n^{-1+D\bar{m}}\exp\left(\left(\bar{g}-D\right)n-\frac{n^{2}}{2}\right)\,,
\]

\noindent with $Z$ the normalization factor. The mean population
size $\bar{m}$ is then obtained from the self-consistency condition
$\bar{m}=\chi_{\bar{g}}(\bar{m})$ where
\begin{equation}
\chi_{\bar{g}}(\bar{m})=\left[\int_{0}^{+\infty}dn\,n^{D\bar{n}}\exp\left(\left(\bar{g}-D\right)n-\frac{n^{2}}{2}\right)\right]/\left[\int_{0}^{+\infty}dn\,n^{-1+D\bar{m}}\exp\left(\left(\bar{g}-D\right)n-\frac{n^{2}}{2}\right)\right]\,.\label{eq:self_cons}
\end{equation}

\noindent We are interested in finding $\bar{g}_{c}$, the minimal
value of $\bar{g}$ such that there exists a positive solution to
Eq.~(\ref{eq:self_cons}). Expanding the self-consistency condition
at small $\bar{m}\ll1$ yields $\bar{m}\simeq\bar{m}\chi'_{\bar{g}}(0)+\bar{m}^{2}\chi''_{\bar{g}}(0)/2$,
which leads to the equation satisfied by the critical growth rate
\begin{equation}
\chi'_{\bar{g}_{c}}(0)=D\int_{0}^{+\infty}dn\,\exp\left(\left(\bar{g}_{c}-D\right)n-\frac{n^{2}}{2}\right)=1\,.\label{eq:critical_point}
\end{equation}
From there, one concludes that $\bar{g}_{c}>0$, with $\bar{g}_{c}\to\infty$
as $D\to0$ and $\bar{g}_{c}\to0$ as $D\to\infty$. Close to the
critical point, one also gets the mean population size $\bar{m}\sim(\bar{g}-\bar{g}_{c})$.
Lastly, taking the average of Eq.~(\ref{eq:DP_fully_connected})
shows that the variance scale in the same way since
\begin{equation}
\left\langle n^{2}\right\rangle =\bar{g}\bar{m}\,.\label{eq:scaling_relation}
\end{equation}

\noindent Beyond the static properties derived in \citep{ottino2020population}, it is of interest to the present
discussion to consider the two-time correlation function $C_{\bar{g}}(t)\equiv\left\langle \left(n(t)-\bar{m}\right)\left(n(0)-\bar{m}\right)\right\rangle $
which is expressed in terms of the transition probability $P_{\bar{g}}[n,t|n_{0}]$
associated to Eq.~(\ref{eq:DP_fully_connected}) as
\begin{equation}
C_{\bar{g}}(t)=\int dndn_{0}\,P_{\bar{g}}(n_{0})P_{\bar{g}}[n,t|n_{0}]\,\left(n-\bar{m}\right)\left(n_{0}-\bar{m}\right)\,.\label{eq:correl_def}
\end{equation}

\noindent Crucially, the transition probability does not exhibit any
sign of critical slowing down as $\bar{g}\to\bar{g}_{c}$. For instance,
for $\bar{g}-D<0$ and neglecting the quadratic term, its expression
is known \citep{feller1951two}, and becomes at large times 
\begin{equation}
P_{\bar{g}}[n,t|n_{0}]\simeq P_{\bar{g}}(n)\left[1+\left(\frac{\left(D-\bar{g}\right)^{2}}{D}\frac{n_{0}n}{\bar{m}}-\left(D-\bar{g}\right)n+D\bar{m}\right){\rm e}^{-(D-\bar{g})t}+O({\rm e}^{-2(D-\bar{g})t})\right]\,.\label{eq:expression_propagator}
\end{equation}

\begin{figure}[h!]
\begin{centering}
\includegraphics[scale=0.5]{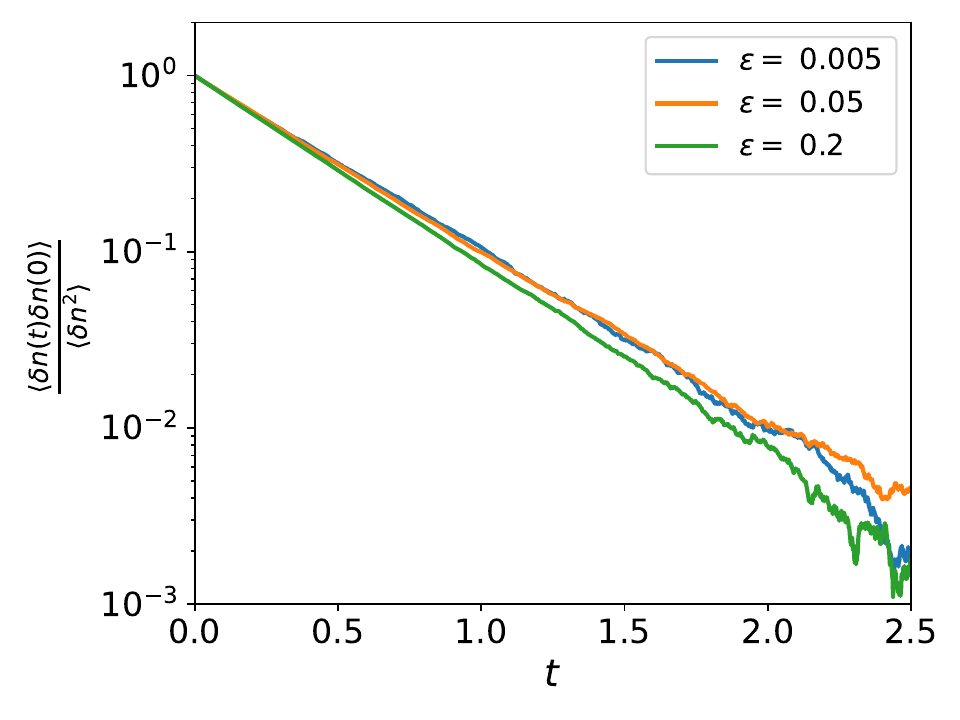}\includegraphics[scale=0.5]{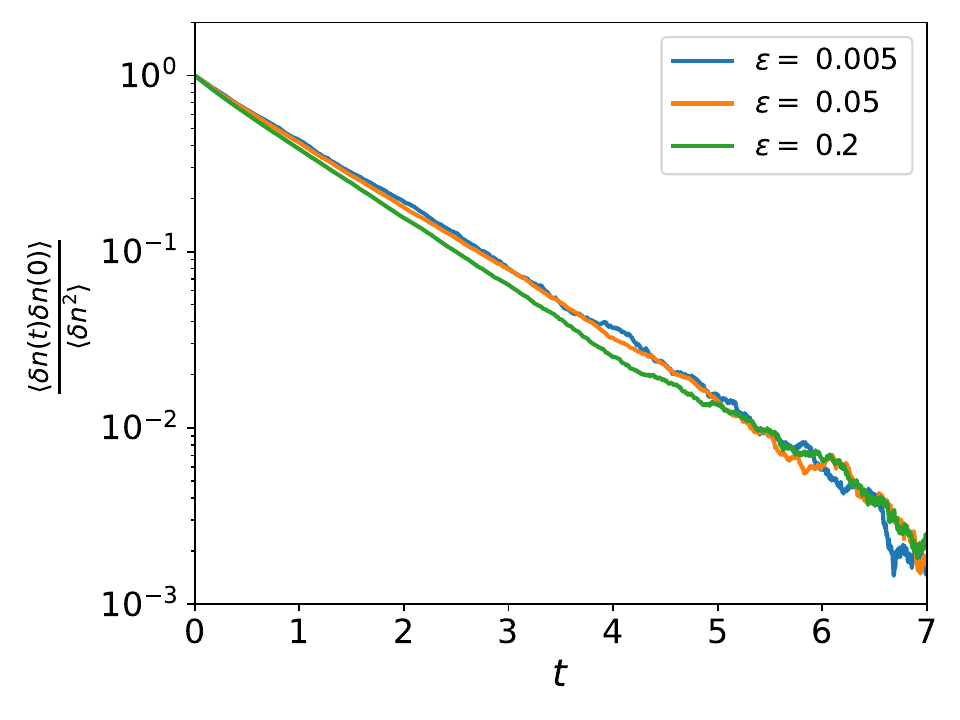}
\par\end{centering}

\caption{Two-time correlation function of the dynamics in Eq.~(\ref{eq:DP_fully_connected}), with $\delta n(t) = n(t) - \bar{m}$. As the critical point is approached $\epsilon=\bar{g}-\bar{g}_{c}\to0$,
the correlation function, rescaled by its value at equal time, converges
to an exponentially decaying function with an $\epsilon$-independent
characteristic timescale. This holds for $D-\bar{g}_{c}<0$ (left)
and for $D-\bar{g}_{c}>0$ (right). We have followed \citep{dornic2005integration,weissmann2018simulation}
for simulating the demographic noise. Parameters: $D=2$ (left) and
$D=0.5$ (right), $dt=10^{-3}$. }
\label{fig:correl}
\end{figure}

\noindent This yields the late-time decay of the correlation function
as 
\begin{align*}
C_{\bar{g}}(t) & ={\rm e}^{-(D-\bar{g})t}\left[\frac{\left(D-\bar{g}\right)^{2}}{D}\frac{\left\langle n^{2}\right\rangle _{\bar{g}}^{2}}{\left\langle n\right\rangle _{\bar{g}}}-\left(D-\bar{g}\right)\left\langle n^{2}\right\rangle _{\bar{g}}\left\langle n\right\rangle _{\bar{g}}+\frac{D}{T}\left\langle n\right\rangle _{\bar{g}}^{3}\right]+O({\rm e}^{-2(D-\bar{g})t})\,,\\
 & \simeq\bar{m}\,{\rm e}^{-(D-\bar{g}_{c})t}\left[\frac{\left(D-\bar{g}_{c}\right)^{2}}{D}\bar{g}_{c}^{2}\right]+O({\rm e}^{-2(D-\bar{g}_{c})t})\,,
\end{align*}
where the last line was obtained to leading order in $\bar{g}-\bar{g}_{c}$.
This expression (which is valid only for $D-\bar{g}_{c} \gg 1$ so that the quadratic term can be safely ignored) shows that, to leading
order in the system size $M$, the decay rate of correlations remains
finite as $\bar{g}\to\bar{g}_{c}$. We have verified numerically for
different values of $D$ (for which $D-\bar{g}_{c}$ can be positive
or negative) that 
\[
\bar{m}^{-1}C_{\bar{g}}(t)\underset{\bar{g}\to\bar{g}_{c}}{\to}h(t)\,,
\]

\noindent with $h(t)$ an exponentially decaying function at large
times, see Fig.~\ref{fig:correl}. To leading order in the system
size, the integrated population fluctuations are thus of the form
\[
\int d\bar{g}\,P(\bar{g})\left\langle \left(n_{\mu}(t)-\left\langle m\right\rangle _{\bar{g}}\right)\left(n_{\nu}(t')-\left\langle m\right\rangle _{\bar{g}}\right)\right\rangle _{\bar{g}}=\delta_{\mu\nu}H(t-t')\,,
\]

\noindent where $H(t)$ decays exponentially at large times.
Therefore, to leading order in the system size, we do not expect the
solution of Eq.~(\ref{eq:self_consistency_fully_connected}) to exhibit power-law
tails.

\subsubsection*{Gaussian corrections}

The large deviation form in Eq.~(\ref{eq:action_DP_fully_connected})
shows that in the active phase, for $\bar{g}>\bar{g}_{c}$, a global
extinction in finite time is exponentially unlikely as $M\to\infty$.
This allows us to study the finite-size fluctuations of $m(t)$ in
the active phase, to first order in $1/\sqrt{M}$. We expand the action
in Eq.~(\ref{eq:action_DP_fully_connected}) around the steady-state
saddle-point solution, with $\hat{m}(s)\to\delta\hat{m}(s)/\sqrt{M}$
and $m(s)\to\bar{m}+\delta m(s)/\sqrt{M}$, and obtain
\begin{align}
M\mathcal{S}[\hat{m},m]= & -i\int dsds'\,\delta\hat{m}(s)\left[\delta(s-s')-DR_{\bar{g}}(s-s')\right]\delta m(s') \nonumber \\ & +\frac{1}{2}\int dsds'\,\delta\hat{m}(s)\delta\hat{m}(s')\left\langle \left(n(s)-\bar{m}\right)\left(n(s')-\bar{m}\right)\right\rangle +O\left(M^{-1/2}\right)\,,
\end{align}

\noindent where $R_{\bar{g}}(t)$ is the response function 
\[
R_{\bar{g}}(t)=\frac{\delta\left\langle n\right\rangle (t)}{\delta h(0)}\,,
\]

\noindent to a perturbation of the form 
\[
\dot{n}=n(\bar{g}-n)+D(\bar{m}-n)+\sqrt{2n}\eta(t)+h(t)\,.
\]

\noindent In Fourier space, the correlations of the Gaussian fluctuations
$\delta m$ are thus given by $\left\langle \delta m(\omega)\delta m(\omega')\right\rangle =2\pi\delta(\omega+\omega')\hat{G}_{\bar{g}}(\omega)$
with
\begin{equation}
\hat{G}_{\bar{g}}(\omega)=\frac{\hat{C}_{\bar{g}}(\omega)}{(1-D\hat{R}_{\bar{g}}(\omega))(1-D\hat{R}_{\bar{g}}(-\omega))}\,,\label{eq:solution_gaussian_correl}
\end{equation}

\noindent where the function $C_{\bar{g}}(t)$ was introduced in Eq.~(\ref{eq:correl_def}). The above expression entails a diverging correlation
time for the fluctuations close to the transition as $D\hat{R}_{\bar{g}_{c}}(0)=1$
at the critical point. Indeed, $\hat{R}_{\bar{g}}(0)$ is given by
the static response function 
\[
\hat{R}_{\bar{g}}(0)=\partial_{h}\left\langle n\right\rangle _{h=0}\,,
\]

\noindent in the presence of a constant external field $h$. Using
the notations introduced in Eq.~(\ref{eq:self_cons}), we have
\[
\left\langle n\right\rangle _{h}=\chi_{\bar{g}}\left(\bar{m}+hD^{-1}\right)\,,
\]

\noindent so that
\begin{equation}
D\hat{R}_{\bar{g}}(0)=\chi_{\bar{g}}'(\bar{m})\,.\label{eq:static_response}
\end{equation}

\noindent We have shown in Eq.~(\ref{eq:critical_point}) that $\chi_{\bar{g}_{c}}'(0)=1$.
Hence, as $\bar{g}\to\bar{g}_{c}$, we get $1-D\hat{R}_{\bar{g}}(0)\sim\bar{g}-\bar{g}_{c}$.
It is further interesting to note that, as $C_{\bar{g}}(t)$, the
response function $R_{\bar{g}}(t)$ does not exhibit any form of critical
slowing down as $\bar{g}\to\bar{g}_{c}$. In fact, we have the
fluctuation-dissipation theorem
\[
R_{\bar{g}}(t)=-\theta(t)\partial_{t}\left\langle n(t)\ln n(0)\right\rangle \,.
\]

\noindent from which we get $R_{\bar{g}}(0{{}^+})=1$ by considering
the dynamics of $f(t)=n(t)\ln n(t)-n(t)$. More generally, the expression
of the propagator given in Eq.~(\ref{eq:expression_propagator})
shows that there is no slow timescale in the response function as
the critical point is approached, and that it remains of $O(1)$ even
as $\bar{m}\to0$ since $\left|\left\langle \ln n\right\rangle \right|\sim m^{-1}$. 

\medskip{}

\noindent At small $\omega$ and close to $g_{c}$ , we can thus approximate
the correlation function of the Gaussian fluctuations $\delta m(t)$
as
\[
\hat{G}_{\bar{g}}(\omega)\simeq\frac{\hat{C}_{\bar{g}}(0)}{\left[\left(1-D\hat{R}_{\bar{g}}(0)\right)^{2}+D^{2}\omega^{2}|\hat{R}'_{\bar{g}_{c}}(0)|^{2}\right]}\,.
\]

\begin{figure}[h!]
\begin{centering}
\includegraphics[scale=0.5]{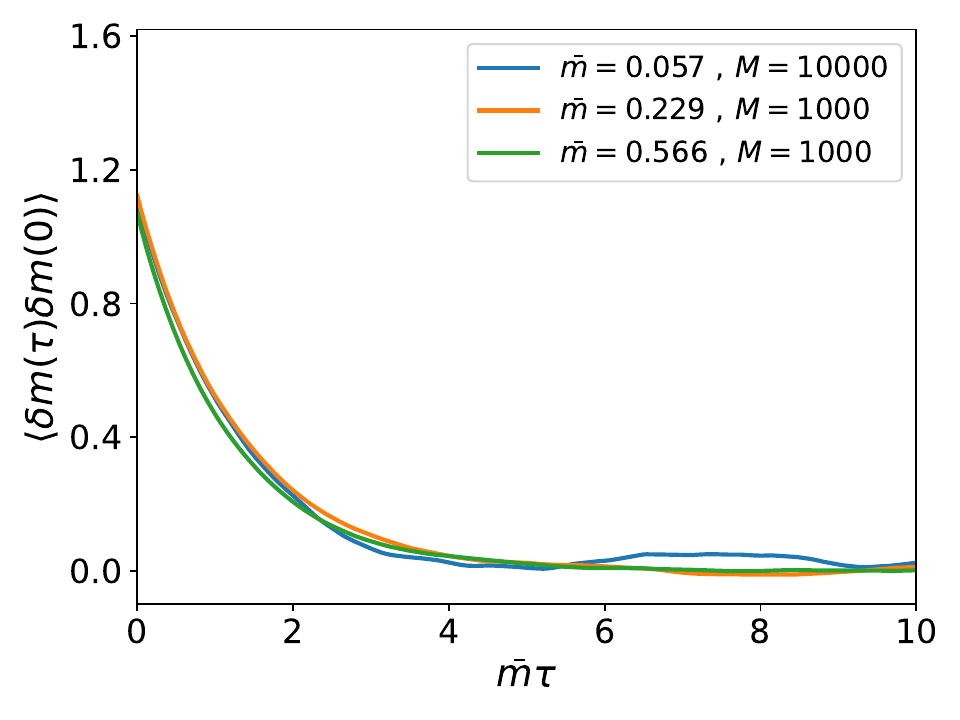}
\par\end{centering}

\caption{Two-time correlation function of the fluctuations $\delta m(t)=\sqrt{M}(m(t)-\bar{m})$
of the total population size for different system sizes $M$ and mean
population sizes $\bar{m}$. As the critical point is approached $\bar{g}\to\bar{g}_{c}$
and $\bar{m}\to0$, the correlation time grows as $1/\bar{m}$, but
the amplitude of the fluctuations remains constant. Parameters: $D=2$,
$dt=10^{-3}$. The dynamics were initialized with $n_{\mu}(t)=\bar{m}$
to ensure a rapid equilibration during a time window $T_{{\rm init}}=100$,
and then run for a time $T=20\,000$ to collect statistics. We have
followed \citep{dornic2005integration,weissmann2018simulation} for
simulating the demographic noise. }
\label{fig:correl_DP}
\end{figure}

\noindent In real-time, and in the vicinity of the critical point,
this gives the long-time decay
\begin{equation}
G_{\bar{g}}(\tau)\sim\exp\left(-\frac{\left|1-D\hat{R}_{\bar{g}}(0)\right|}{D|\hat{R}'_{\bar{g}_{c}}(0)|}\tau\right)\,,\label{eq:correl_large_scale}
\end{equation}

\noindent with a correlation time that grows as $|\bar{g}-\bar{g}_{c}|^{-1}$
and an amplitude that remains of order $O(1)$ even as $\bar{m}\to0$
as both $\hat{C}_{\bar{g}}(0)\sim\bar{g}-\bar{g}_{c}$ and $1-D\hat{R}_{\bar{g}}(0)\sim\bar{g}-\bar{g}_{c}$, see Fig.~\ref{fig:correl_DP}. The resulting
fluctuations are thus much larger than suggested by a naive estimate
from independent sites evolving according to Eq.~(\ref{eq:DP_fully_connected_many_body}),
\[
\frac{1}{M^{2}}\sum_{\mu}\left\langle \left(n-\left\langle n\right\rangle \right)^{2}\right\rangle \sim\frac{\bar{m}}{M}\,.
\]

\subsubsection*{Large-scale behavior}

\noindent To order $1/\sqrt{M}$, the populations at the different
sites evolve according to 
\[
\dot{n}_{\mu}=n_{\mu}(\bar{g}-n_{\mu})+D\left(\bar{m}+\frac{1}{\sqrt{M}}\delta m(t)-n_{\mu}\right)+\sqrt{2n_{\mu}}\eta_{\mu}(t)\,,
\]
where $\delta m(t)$ is a Gaussian process with correlations given
in Eq.~(\ref{eq:solution_gaussian_correl}). The large-scale form
in Eq.~(\ref{eq:correl_large_scale}) therefore shows that the integrated
fluctuations of population sizes decay as a power-law with an amplitude
that decreases with the system size
\begin{equation}
\int d\bar{g}\,P(\bar{g})\left\langle \left(n_{\mu}(t)-\left\langle m\right\rangle _{\bar{g}}\right)\left(n_{\nu}(t')-\left\langle m\right\rangle _{\bar{g}}\right)\right\rangle _{\bar{g}}\sim\frac{A_{M}}{\left|t-t'\right|}\,,\label{eq:DP_criticality}
\end{equation}
where $A_{M}\sim M^{-1}$, and which holds whether $\mu=\nu$ or $\mu\neq\nu$.
Thus, the solution of Eq.~(\ref{eq:self_consistency_fully_connected}) must exhibit
power-law tails, with an amplitude that decreases with the system
size. As the result derived in Eq.~(\ref{eq:DP_criticality}) is
based on the near-critical properties of the directed percolation
theory on the fully connected lattice, which are left unchanged in the presence of short-ranged environmental noise
\citep{ottino2020population}, we conjecture that Eq.~(\ref{eq:DP_criticality})
provides the correct scaling for the large-scale behavior of the solution
of the dynamical mean-field theory equation, with possible logarithmic
corrections. In pratice however, the subleading Gaussian fluctuations
investigated here are large enough to provoke finite time extinctions
of the near-critical modes, that is for $\sqrt{M}\bar{m}\lesssim1$.
Large-scale fluctuations in finite size systems are therefore expected
to behave as 
\[
\int d\bar{g}\,P(\bar{g})\left\langle \left(n_{\mu}(t)-\left\langle m\right\rangle _{\bar{g}}\right)\left(n_{\nu}(t')-\left\langle m\right\rangle _{\bar{g}}\right)\right\rangle _{\bar{g}}\sim\frac{M^{-1}}{\left|t-t'\right|}\exp\left(-\alpha\left|t-t'\right|/\sqrt{M}\right)\,,
\]
for some positive number $\alpha$.

\end{document}